\documentclass[twocolumn,
trackchanges]{aastex7}
\usepackage{soul}
\usepackage{CJK}
\usepackage{subfigure,soul}
\usepackage[normalem]{ulem}
\usepackage{graphicx}
\usepackage{dcolumn}
\usepackage{bm}
\usepackage{mathrsfs}
\usepackage{amsthm,amsmath,amssymb}
\usepackage{braket}
\usepackage{graphicx}
\usepackage{xcolor}  
\usepackage{multirow}
\usepackage{footmisc}

\definecolor{deepblue}{RGB}{255,0,0}

\accepted{September 8, 2025}
\submitjournal{ApJ}

\begin{document}
\begin{CJK*}{UTF8}{gbsn}
\title{Tidal disruption events in active galactic nuclei: on orbital inclination and Schwarzschild apsidal precession 
}

\correspondingauthor{Wenda Zhang}
\author[0009-0006-5706-0364]{Minghao Zhang (张铭浩)}
\affiliation{National Astronomical Observatories, Chinese Academy of Sciences\\
 Beijing 100012 People's Republic of China; \url{zhangmh@bao.ac.cn}, \url{wdzhang@nao.cas.cn}} 
\affiliation{School of Astronomy and Space Science, University of Chinese Academy of Sciences, Beijing 100049, People's Republic of China}
\email{zhangmh@bao.ac.cn}

\author[0000-0003-1702-4917]{Wenda Zhang (张文达)}
\affiliation{National Astronomical Observatories, Chinese Academy of Sciences\\
 Beijing 100012 People's Republic of China; \url{zhangmh@bao.ac.cn}, \url{wdzhang@nao.cas.cn}} 
\email{wdzhang@nao.cas.cn}  

\author[0000-0003-1702-4917]{Hongping Deng (邓洪平)}
\affiliation{Shanghai Astronomical Observatory, Chinese Academy of Sciences, 80 Nandan Road, Shanghai 200030, People's Republic of China}
\email{hpdeng353@shao.ac.cn}  

\author[0000-0001-8416-7059]{Hengxiao Guo (郭恒潇)}
\affiliation{Shanghai Astronomical Observatory, Chinese Academy of Sciences, 80 Nandan Road, Shanghai 200030, People's Republic of China}
\email{hengxiaoguo@gmail.com}  

\author[0000-0003-3472-4392]{Jingbo Sun (孙静泊)}
\affiliation{Shanghai Astronomical Observatory, Chinese Academy of Sciences, 80 Nandan Road, Shanghai 200030, People's Republic of China}
\affiliation{School of Astronomy and Space Science, University of Chinese Academy of Sciences, Beijing 100049, People's Republic of China}
\email{sunjingbo@shao.ac.cn}  

\begin{abstract}
Tidal disruption events (TDEs) in active galactic nuclei (AGNs) mark a regime where traditional vacuum models fail to capture the full dynamics, especially due to interaction between stellar debris and pre-existing accretion disks. We perform meshless hydrodynamic simulations incorporating both general relativistic (GR) effects and radiative cooling to study TDEs in AGNs with different orbital inclinations ($\theta_{\rm inc}$) of the disrupted star, ranging from projected prograde to retrograde orbits.
We post-process the simulations to derive multi-wavelength light curves and identify several distinct features in the light curves, including a precursor flare from early debris-disk collision and a major flare driven by fallback.
The dynamics of the stellar debris and accretion disk, and subsequently the light curve features, are strongly affected by $\theta_{\rm inc}$ and GR effects.
Retrograde orbits ($\theta_{\rm inc}=135^\circ$) yield a more luminous, shorter major flare and a more prominent precursor than prograde ones ($\theta_{\rm inc}=22.5^\circ$).
During fallback, prograde cases ($\theta_{\rm inc} = 22.5^\circ$, $45^\circ$) develop a central cavity with spirals in the inner region of the AGN disk, leading to transient UV/X-ray suppression accompanied by oscillations, while higher inclinations ($\theta_{\rm inc}=90^\circ$, $135^\circ$) form a gradually tilting inner disk, potentially causing UV/X-ray dips via geometric effects at certain viewing angles.
Relativistic apsidal precession alters stream collisions, producing structural differences in the inner disk, outer disk, and debris compared to Newtonian cases, and drives quasi-periodic signals in prograde configurations.
These results provide predictive diagnostics for identifying AGN TDEs and interpreting observed light-curve diversity.
\end{abstract}


\section{Introduction}\label{sec:intro}
A tidal disruption event (TDE) occurs when a star enters the tidal radius of a supermassive black hole (SMBH), leading to its destruction by tidal forces. 
During this process, part of the star is torn apart, with some fragments falling back towards the black hole in high-eccentricity orbits \citep{1989ApJ...346L..13E}.
Traditionally, models of TDEs have focused on those occurring in quiescent galaxies, in which the central SMBHs are nearly in vacuum (e.g., \citealt{2013ApJ...767...25G}, \citealt{2016MNRAS.455.2253B}). 
However, TDEs can also occur in an active galactic nucleus (AGN) powered by an accretion disk (e.g., \citealt{2019ApJ...881..113C}, \citealt{McKernan_2022}, \citealt{2024ApJ...962L...7W}, \citealt{2024ApJ...977...63L}).

Theoretically, the presence of an accretion disk could enhance TDE rates through disk-star interactions (e.g., \citealt{2012ApJ...758...51J}, \citealt{2016MNRAS.460..240K}, \citealt{2024ApJ...962L...7W}).
It is indicated that TDE rate can be amplified by several orders of magnitude during the AGN phase than in the quiescent state through disk capture \citep{2024ApJ...962L...7W}. 
Observationally, several AGN TDE candidates (e.g., \citealt{Bao_2024}, \citealt{Jiang_2019}, \citealt{2025ApJ...982..150S}, \citealt{2019ApJ...883...94T}, \citealt{Payne_2021}) have been discovered through large sky surveys such as All-Sky Automated Survey for SuperNovae (ASAS-SN, \citealt{Shappee_2014}), Asteroid Terrestrial-impact Last Alert System (ATLAS, \citealt{2018PASP..130f4505T}) and Zwicky Transient Facility (ZTF, \citealt{2019PASP..131a8002B}). 
Beyond optical and ultraviolet (UV) flares, some exhibit distinctive multi-wavelength variability, particularly anti-correlated X-ray suppression (e.g., AT2021aeuk; \citealt{2025ApJ...982..150S}, 1ES 1927+654; \citealt{2019ApJ...883...94T}, ASASSN-14ko; \citealt{Payne_2021}, AT2019aalc; \citealt{2024arXiv240817419V}). 
It is worth noticing that a peculiar UV variability pattern during the X-ray suppression has also been observed in the second flare of AT2021aeuk, where the UV flux first rises, then declines below its initial level, and eventually returns to the pre-flare level \citep{2025ApJ...982..150S}.
Similar late-time UV suppression is also seen in other AGN TDE candidates (e.g., \citealt{Cannizzaro_2022}), supporting the scenario where the AGN disk is partially disturbed or disrupted by TDE.
Substantial diversity among AGN TDE candidates suggests more intricate underlying processes, highlighting the necessity of exploring a broader parameter space in numerical simulations.

Hydrodynamic simulations investigating the impact of stream collisions on both the debris structure, the AGN disk, and the dusty torus are conducted using various numerical methods (e.g., \citealt{2019ApJ...881..113C}, \citealt{2024MNRAS.527.8103R}, \citealt{2024ApJ...977...63L}).
Through moving-mesh hydrodynamic simulations, \cite{2024MNRAS.527.8103R} demonstrates that due to continuous interaction with a pre-existing accretion disk, debris of in-disk TDEs exhibits substantial deviations in orbital energy distribution ($dM/dE$) from that of naked TDEs.
As a result, the light curves of AGN TDEs cannot be simply modeled as a superposition of a vacuum TDE and intrinsic AGN emission.
Furthermore, the fallback debris can strongly impact the AGN structure itself.
Through grid-based hydrodynamic simulations, \cite{2019ApJ...881..113C} found that when debris collides with the inner disk, it can trigger spiral shocks, removing angular momentum and enhancing the inflow rate by orders of magnitude. 
Additionally, the mixing of matter from stellar debris into the AGN disk can enhance its metallicity, providing a mechanism distinct from supernova enrichment \citep{2024MNRAS.527.8103R}.
Beyond the accretion disk, the geometrically thick dusty torus could also be affected by TDE debris. \cite{2024ApJ...977...63L} reveals that the collision between the unbound TDE debris and the dusty torus could be a potential mechanism for the delayed radio outbursts.

For simulations focusing on debris-disk interaction, in-plane TDEs with prograde and retrograde stellar orbits \citep{2024MNRAS.527.8103R} and off-plane TDEs with orbits perpendicular to the disk \citep{2019ApJ...881..113C} have been studied. However, the role of the star's orbital inclination in shaping the resulting accretion flow remains unexplored.
Due to conservation of angular momentum, the collision between the fallback stream and the disk may lead to either the formation of a tilted inner disk or the destruction of the region inside the impact radius. 
These diverse inner disk structures could in turn produce distinct radiative signatures, potentially explaining the observed diversity in the multi-band light curves of AGN TDEs.
To address this gap, we aim to investigate how orbital inclinations of disrupted stars affect the formation and long-term evolution of the inner accretion disk while attempting to produce the multi-band radiative features based on the simulated data.

In addition, general relativistic (GR) effects have been shown to be non-negligible for TDEs of main-sequence (MS) stars (e.g., \citealt{2016MNRAS.455.2253B}, \citealt{2023ApJ...957...12R}, \citealt{2024ApJ...971L..46P}, \citealt{2016MNRAS.461.3760H}). 
Due to strong apsidal precession, the stream can be self-intersected while falling back, accelerating the circularization process \citep{2016MNRAS.455.2253B}.
For in-disk TDEs, \cite{2024MNRAS.527.8103R} pointed out that the perturbation of the inner AGN disk at the first pericenter passage would be stronger for more relativistic cases.
Additionally, it is expected that the deviation of GR trajectories from the Newtonian case will introduce additional discrepancies in the continuous interaction between the debris and the disk. 
This will alter the shape of the debris and the distribution of its orbital energy, ultimately affecting the fallback process, leading to markedly different disk structures. 
Furthermore, several other phenomena potentially linked to AGN TDEs may involve GR effects, such as quasi-periodic eruptions (QPEs) and quasi-periodic oscillations (QPOs).
The origin of QPEs is still uncertain, but some studies point to a possible connection with TDEs and AGN activity, based on observations showing that some QPEs occur after TDEs (e.g. \citealt{2024Natur.634..804N}, \citealt{2023A&A...675A.152Q}), the presence of AGN signatures in their host galaxies \citep{Wevers_2022}, and similarities between the host galaxies of QPEs and TDEs \citep{Jiang_and_Pan_2025}.
Moreover, QPOs that are commonly seen in X-ray binaries \citep{QPO_review} have also been reported in AGNs (e.g.,  \citealt{2025Natur.638..370M}, \citealt{Pan_2016}).
These quasi-periodic behaviors are suggested to be driven by GR apsidal or nodal precession of asymmetric accretion flow (e.g. \citealt{1999ApJ...524L..63S}, \citealt{Raj_2021a}, \citealt{Raj_2021b}).
Therefore, through numerical simulations, we aim to provide a detailed depiction of how GR apsidal precession affects the debris-disk interaction, and to explore whether debris-disk collisions in AGN TDEs in a Schwarzschild gravitational field can excite quasi-periodic behaviors in the accretion flow.

The paper is structured as follows.
In Section \ref{sec:numerical}, we present the numerical setup, including the initial conditions of the star and disk, the treatment of radiative cooling, and the implementation of GR corrections.
Section \ref{sec:numerical_results} discusses the outcome of our parameter exploration, including the effect of relativistic apsidal precession and the dependence on stellar orbital inclination. 
Synthetic multi-band light curves are derived from simulated data and discussed in Section \ref{sec:Synthetic}.
Discussions and applications to observations based on our numerical results are in Section \ref{sec:Discussions}.

\section{Numerical Setup}\label{sec:numerical}
We simulated a self-gravitating star tidally disrupted by a SMBH of mass 
$M_{\rm BH}=10^6M_\odot$, surrounded by a thin accretion disk.  
To efficiently calculate self-gravity and evolve the multi-scale physical processes involved in TDEs in AGNs, we use the Godunov-type Lagrangian hydrodynamics algorithm, Meshless Finite Mass (MFM), as implemented in the GIZMO code \citep{2015MNRAS.450...53H}. 
In contrast to Smoothed Particle Hydrodynamics (SPH) codes, 
MFM resolves shocks within the Godunov framework rather than introducing an artificial viscosity, leading to improved accuracy in subsonic flow behavior and precise shock resolution (e.g., \citealt{Deng_2019}).
Compared to Eulerian grid-based methods, MFM significantly reduces advection errors and directional biases caused by fixed mesh orientation \citep{2015MNRAS.450...53H}.

\subsection{Viscosity of the Accreting System}
To introduce physical viscosity for an accreting system, the standard Navier-Stokes equations are solved in GIZMO \citep{2017MNRAS.466.3387H}, in which the viscous stress tensor reads  
\begin{equation}
\mathcal{T}_{\text{vis},ij}=-\eta\left(\frac{\partial v_i}{\partial x_j}+\frac{\partial v_j}{\partial x_i}-\frac{2}{3}\frac{\partial v_k}{\partial x_k}\delta_{ij}\right),
\end{equation}
where $\eta=\rho\nu$ is the dynamic viscosity that can be given to an arbitrary value in the code, and $\nu=\alpha c_sH$ is the kinematic viscosity. Here the parameterized $\alpha$ prescription of viscosity is adopted \citep{1973A&A....24..337S}.
For a disk with given density $\rho$ and aspect ratio $H/R$, the corresponding $\eta$ for a given $\alpha$ is
\begin{equation}
\eta=\rho\alpha c_sH \approx\rho\alpha \left(\frac{H}{R}\right)^2Rv_k.
\end{equation}
In all our models, a typical value $\alpha=0.1$ for AGN disks (e.g., \citealt{King_2007}, \citealt{Liu_2008}, \citealt{Martin_2019}) is adopted.

\subsection{Gravitational Potential of the Central Black Hole}
Previous studies have demonstrated that relativistic effects can play a crucial role in TDEs. 
For instance, apsidal precession can cause the fallback material to intersect, leading to strong shocks that facilitate rapid circularization of the debris and disk formation (e.g., \citealt{2016MNRAS.455.2253B}, \citealt{2023ApJ...957...12R}).
However, solving the full GR equations in hydrodynamic simulations could be complex and computationally expensive (but see \citealt{Fedrigo_Lupi_2025}). 
To address this, \cite{2013MNRAS.433.1930T} developed a generalized Newtonian potential by solving the geodesic equation for a time-like particle in the Schwarzschild metric. 
This approach allows us to easily incorporate the effect of GR apsidal precession into Newtonian hydrodynamics codes with minimal additional computational cost.
The generalized Newtonian potential (hereafter the GR potential) reads
\begin{equation}
\Phi_{\rm G}=-\frac{GM_{\rm BH}}{R}-\left(\frac{2R_{\rm g}}{R-2R_{\rm g}}\right)\left[\left(\frac{R-R_{\rm g}}{R-2R_{\rm g}}\right)v_r^2+\frac{v_t^2}{2}\right],
\end{equation}
where $R_g=\frac{GM_{\rm BH}}{c^2}$ is the gravitational radius of the black hole. Beyond radial dependence, this potential is also a function of local velocity, with radial and tangential components denoted as $v_r$ and $v_t$, respectively. 
To highlight the effects of GR apsidal precession, we also carried out simulations 
with Newtonian potential
\begin{equation}
\Phi_{\rm N}=-\frac{GM_{\rm BH}}{R}
\end{equation}
for comparison.

In both gravitational potentials, a spherical cutout of radius $R_{\rm acc}=6R_{\rm g}$ is set as the accretion radius, which corresponds to the innermost stable circular orbit (ISCO) of a Schwarzschild black hole. 
Particles entering this radius will be cleared.

\subsection{Radiative Cooling Term}\label{sec:rad_cooling}
Previous numerical simulations of AGN TDEs have typically adopted an adiabatic equation of state (EOS) (e.g., \citealt{2019ApJ...881..113C}, \citealt{2024MNRAS.527.8103R}). 
While the adiabatic process is a good approximation during the disruption phase for vacuum TDEs (e.g., \citealt{2013ApJ...767...25G}), it becomes inadequate for modeling the debris-disk collision in AGN environments, where radiative cooling can play a significant role.

Insights from vacuum TDE simulations suggest that the disk formation process following stream collisions is highly sensitive to the radiative cooling efficiency.
Specifically, the fall-back stream tends to form a geometrically thick torus in adiabatic models representing the radiatively inefficient limit, while a geometrically thin ring structure forms under isothermal or isentropic EOS that stands for radiatively efficient cases (e.g., \citealt{2016MNRAS.455.2253B}, \citealt{Liptai_2019}, \citealt{2024ApJ...971L..46P}).
Furthermore, in simulations of accretion disks, radiative cooling efficiency is proved to have strong impact on the formation and evolution of substructures of the disk, such as spiral shocks, cavities, and gaps (e.g., \citealt{2020MNRAS.499.2836F}, \citealt{2024A&A...688A.174F}, \citealt{2024ApJ...961...86Z}).
These considerations motivate the inclusion of radiative cooling in AGN TDE simulations to more accurately capture the hydrodynamics of stream collision and the resulting disk morphology.

In the absence of a fully self-consistent radiative transfer treatment, we incorporate an approximate radiative cooling term to evolve the specific internal energy of the fluid. 
This term promptly removes excess energy generated by shock heating during stream collisions and viscous dissipation within the disk.
For computational efficiency, we estimate the net radiative energy loss of each fluid element $i$ from local quantities (e.g., \citealt{Wilkins_Clarke_2012}; \citealt{2015MNRAS.447...25L}; \citealt{Young_2024}).
The temperature $T_i$ is computed from the specific internal energy $u_i$ via the ideal gas law with $\gamma = 5/3$:
\begin{equation}
P_i = (\gamma - 1) \rho_i u_i = \frac{\rho_i}{\mu m_p} k_B T_i,
\end{equation}
where $P_i$ is the gas pressure, $\rho_i$ is the density, $\mu=0.6$ is the mean molecular weight, $m_p$ is the proton mass, and $k_B$ is the Boltzmann constant.
We assume that the emissivity of each fluid element follows a blackbody form, $\sigma T_i^4$, where $\sigma$ is the Stefan-Boltzmann constant.

To model the overall effect of radiative self-absorption, we apply a diffusion-like flux model, 
\begin{equation}
\left(\frac{d(\rho_i u_i)}{dt}\right)_{\rm cool}=-\frac{dF_i}{dz},
\end{equation} 
with the radiative flux defined as
\begin{equation}
F_i =-\frac{16}{3\kappa\rho_i}\sigma T_i^3\frac{dT_i}{dz}.
\end{equation}
Here we adopt a Rosseland mean opacity
$
\kappa = 5\times10^{24}\rho_i T_i^{-7/2}\text{cm}^2\text{g}^{-1}
$ (e.g., \citealt{2002apa..book.....F}).
Under the thin-disk approximation, the vertical gradient $d/dz$ is replaced by $1/H_i$ \citep{2002apa..book.....F}, with $H_i$ estimated from the pressure gradient \citep{2015MNRAS.447...25L}:
\begin{equation}
H_i \approx \frac{P_i}{\nabla P_i} = \frac{P_i}{\rho_i |\textbf{\emph{a}}_{i,\mathrm{hydro}}|},
\end{equation}
where the hydrodynamic acceleration $\textbf{\emph{a}}_{i,\mathrm{hydro}}$ is directly obtained from the force calculation in the MFM algorithm.
We note that although this diffusion-like approximation may break down near the disk surface, it is valid near the disk midplane which dominates the radiative cooling, and is therefore sufficient for capturing the cooling at the order-of-magnitude level.

Additionally, since the star in our simulation is a self-gravitating polytropic sphere without an internal heat source, unconstrained radiative cooling would disrupt its hydrostatic equilibrium and cause collapse. To prevent this, we impose a cutoff temperature $T_{\rm cut}\approx5\times10^5$\,K, comparable to the stellar core temperature, below which cooling is disabled. This prevents the stellar structure from being altered by the cooling term prior to debris-disk collisions. 
Disk regions with $T<T_{\rm cut}$, corresponding roughly to the outer AGN disk ($R\gtrsim2000R_g$), contribute only $\sim0.2\%$ of the total viscous heating and thus have a negligible effect on the global heating-cooling balance.

\subsection{Initial Conditions: Stellar Structure and AGN Disk Structure}
In our simulations, the star is modeled as a self-gravitating polytropic gas sphere (also adopted in \citealt{2016MNRAS.455.2253B}, \citealt{2016MNRAS.461.3760H}, \citealt{2024ApJ...971L..46P}), with its structure determined by solving the Lane-Emden equation for $\gamma = 5/3$ to maintain a stable structure under our adiabatic EOS before its tidal disruption. 
The mass and radius adopted in all of our main-sequence (MS) star models are $M_\star = 8M_\odot$ and $R_\star = 5R_\odot$, respectively. We choose a stellar mass higher than $1M_\odot$ based on both theoretical predictions and observations (e.g., \citealt{10.1111/j.1365-2966.2006.10772.x}, \citealt{2013A&A...549A..57N}) indicating that massive stars are overrepresented in the Galactic center.
The tidal radius of the star for a $10^6 M_\odot$ SMBH is $R_t=R_\star (M_{\rm BH}/M_\star)^{1/3}\approx 118 R_g$.

To obtain a stable disk structure under radiative cooling, we first select a set of trial temperature and density profiles. 
The trial density profile follows a power-law dependence on radius and an exponential decay in $z$ direction:
\begin{equation}\begin{aligned}
\rho_{\rm disk}(R,Z)=&5\times10^{-7}\text{g/cm}^3\left(\frac{R}{500R_g}\right)^{-p}\\
&\exp\left(
\left(\frac{H}{R}\right)^2
\left(\frac{1}{\sqrt{1+(Z/R)^2}}-1\right)
\right),
\end{aligned}
\end{equation}
where the power-law index is set to
\begin{equation}
p=
\begin{cases}
-1  \  (R<500R_g),\\
\\
-3  \ (R>500R_g).
\end{cases}
\end{equation}
We note that this trial density profile is consistent in form with that in \cite{2024MNRAS.527.8103R} but has a steeper slope in the inner disk.
The trial temperature profile is set according to the ideal-gas equation
\begin{equation}
P=\frac{\rho}{\mu m_p} k_B T
\end{equation}
with pressure estimated by $P=\rho c_s^2\approx\rho(H/R)^2v_k^2$.  
The aspect ratio of the trial disk is set to be $H/R=0.01$ for all models.

\begin{figure}[h]
\centering
\includegraphics[width=0.9\linewidth]{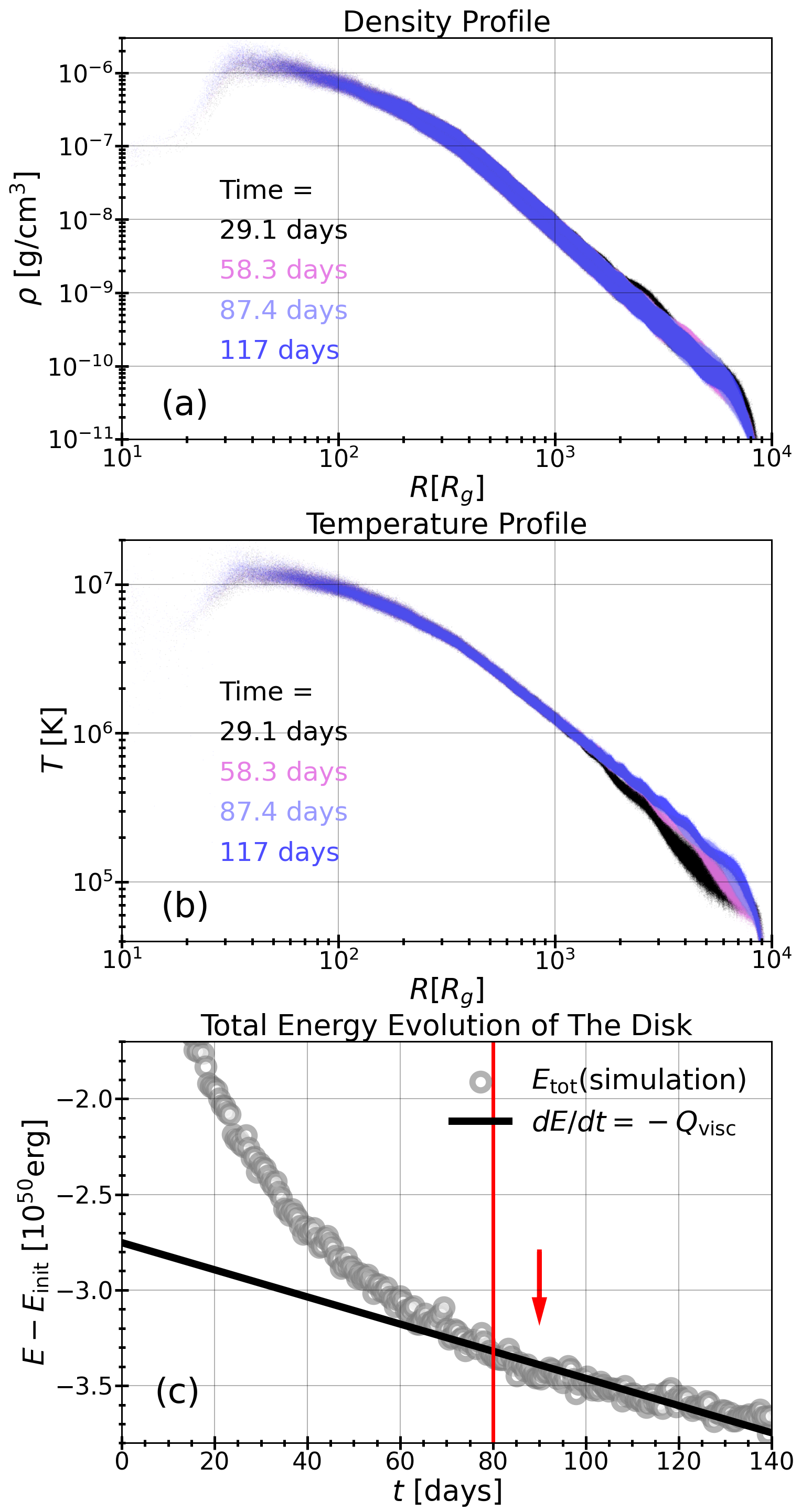}
\caption{
\label{initial_condition} 
Radial profiles of (a) density and (b) temperature of the AGN disk during relaxation, with colors indicating different times.
Panel (c) shows the time evolution of the disk's total energy. Gray circles represent simulation data, while the black solid line indicates the expected quasi-steady energy loss rate, estimated from the viscous heating rate 
$Q_{\rm visc}$ (Equation \ref{viscous_heating}).
The red vertical line marks $t\approx80$ days, when viscous heating becomes nearly balanced by radiative cooling.
The red arrow in panel (c) indicates $t\approx90$ days, at which the disk is considered to be relaxed and the corresponding profiles are adopted as the initial condition.
}
\end{figure}

Next, we relax the trial temperature and density profile under radiative cooling and viscous heating.
As a reference timescale, we adopt the Keplerian orbital period at  $R=10^3R_g$, denoted as $T_{\rm ref}$, which corresponds to $\sim$11.4 days for the adopted $M_{\rm BH}$.
The relaxation process lasts for approximately $12T_{\rm ref}$ ($\sim 140$ days), which is comparable to the timescale of our debris-disk collision experiments.  
Figure \ref{initial_condition}(a) and (b) show the radial profiles of density and temperature during the relaxation process, respectively. 
Both profiles become nearly steady after $\sim2.6T_{\rm ref}$ ($\sim29$ days).

To further examine whether viscous heating is nearly balanced by radiative cooling, we numerically evaluated the energy conservation during the relaxation process by estimating the total heat produced by viscous dissipation ($Q_{\rm visc}$) and carried away from the system via radiative cooling ($(dE_{\rm tot}/dt)_{\rm cool}$).
The former is computed as
\begin{equation}
\label{viscous_heating}
\begin{aligned}
Q_{\rm visc} &\approx \int_{R_{\rm in}}^{R_{\rm out}}\int_{-H(R)}^{H(R)} q_{\rm visc}2\pi RdRdz\\
&\approx 8.2\times10^{42} \text{erg/s},
\end{aligned}
\end{equation}
where
\begin{equation}
q_{\rm visc} = \rho\alpha c_s H\frac{dv_\phi}{dR}
\end{equation}
is the volume rate of viscous dissipation by differential rotation \citep{2002apa..book.....F}.
Here we adopt the density profile at $t\approx8 T_{\rm ref}$ ($\approx90$ days), as it remains nearly unchanged for $t>2.6 T_{\rm ref}$.
Considering that the energy inflow through the accretion radius is negligible compared to radiative cooling once the system reaches a quasi-steady state, the latter can be roughly estimated by the rate of the total energy loss of the disk:
\begin{equation}
\left(\frac{dE_{\rm tot}}{dt}\right)_{\rm cool}\approx\frac{dE_{\rm tot}}{dt}=\frac{d}{dt}\left(E_{\rm orbit}+E_I\right),
\end{equation}
where the orbital energy, internal energy and total energy of the disk are denoted as $E_{\rm orbit}$, $E_I$ and $E_{\rm tot}$, respectively.

Figure \ref{initial_condition}(c) presents the evolution of the system's total energy during the relaxation process. 
For $t>7 T_{\rm ref}$ ($\sim$80 days), the energy loss rate of the system is approximately balanced by the viscous dissipation rate estimated from Equation \ref{viscous_heating}.
Therefore, we adopt the profiles at $t\approx8 T_{\rm ref}$ ($\approx90$ days) as the initial condition for the AGN disk, and further put the star into this environment to investigate its disruption and fallback processes in the vicinity of the disk.

\subsection{Orbital Configurations of the Star}\label{sec:orbit}
\begin{table*}[ht]
\scriptsize
\tabcolsep=0.15cm
\caption{
Parameters of the simulations.
Parameters that are different from the fiducial model are marked in bold.
\label{table:setups}}
\centering
\begin{tabular}{lccccccc}
\hline
Model    & Eccentricity &Impact Parameter&Stellar Mass    & Stellar Radius    &Orbital Inclination   & Gravity & Resolution\\  
         &$e$           &  $\beta$       & $M_\star$      & $R_\star$         & $\theta_{\rm inc}$   &  & $(N_{\rm disk},N_{\rm star})$      \\  
\hline
 \multicolumn{8}{c}{Fiducial model}\\
\hline
R90          & 0.99   & 2       & 8$M_\odot$ & 5$R_\odot$  & 90° &  GR & $(1\times10^6,4\times10^5)$\\
\hline
 \multicolumn{8}{c}{Regular runs}\\
\hline
R22.5      & 0.99   & 2       & 8$M_\odot$ & 5$R_\odot$    & \textbf{22.5°} &  GR &$(1\times10^6,4\times10^5)$\\ 
R45       & 0.99    & 2      & 8$M_\odot$ & 5$R_\odot$     & \textbf{45°} &  GR &$(1\times10^6,4\times10^5)$ \\
N45       & 0.99   & 2      & 8$M_\odot$ & 5$R_\odot$     & \textbf{45°} &  \textbf{Newtonian}&$(1\times10^6,4\times10^5)$\\ 
N90         & 0.99   & 2       & 8$M_\odot$ & 5$R_\odot$    & 90° & \textbf{Newtonian} & $(1\times10^6,4\times10^5)$\\
R135       & 0.99   & 2      & 8$M_\odot$ & 5$R_\odot$   &\textbf{135°} &  GR & $(1\times10^6,4\times10^5)$\\ 
\hline
 \multicolumn{8}{c}{Resolution Tests}\\
\hline
RT45\_0.35M  & 0.99    & 2      & 8$M_\odot$ & 5$R_\odot$ &\textbf{ 45°}  &  GR& 
\textbf{$(2.5\times10^5,1\times10^5)$}\\
RT45\_0.70M  & 0.99  & 2      & 8$M_\odot$ & 5$R_\odot$   &\textbf{ 45°}  &  GR& 
\textbf{$(5\times10^5,2\times10^5)$}\\
RT45\_1.40M  & 0.99  & 2       & 8$M_\odot$ & 5$R_\odot$  &\textbf{ 45°}  &  GR& 
\textbf{$(1\times10^6,4\times10^5)$}\\
RT90\_0.70M  & 0.99   & 2   & 8$M_\odot$ & 5$R_\odot$     &  90°  &  GR &\textbf{$(5\times10^5,2\times10^5)$}\\
RT90\_1.40M  & 0.99   & 2      & 8$M_\odot$ & 5$R_\odot$  &  90°  &  GR &\textbf{$(1\times10^6,4\times10^5)$}\\
\hline
\end{tabular}
\end{table*}

In each of our simulations, the star is initially placed on a highly eccentric bound orbit, with the semi-major axis and eccentricity set to $a=5900R_g$ and $e=0.99$, respectively. 
This corresponds to an impact parameter $\beta=R_t/R_p\approx2$. 
In all cases, the star begins its trajectory at a distance of $7R_t$ from the central SMBH, and its initial velocity is calculated based on the corresponding Keplerian orbital parameters.

The disruption of stars on eccentric orbits is motivated by several astrophysical mechanisms.
For example, a close encounter between a tightly bound binary system and an SMBH can lead to the tidal breakup of the binary, leaving one component bound to the SMBH on a highly eccentric orbit while ejecting the other at high velocity \citep{1988Natur.331..687H}.
In addition, intermediate-mass black holes (IMBHs) and stellar-mass black holes (sBHs) are believed to orbit SMBHs and gradually inspiral due to gravitational wave emission, making them prime targets for next-generation gravitational wave detectors (e.g., Laser Interferometer Space Antenna, LISA; \citealt{LISA_REF}, \citealt{2007CQGra..24R.113A}, \citealt{PhysRevD.95.103012}). 
In such an environment, the capture rate of passing stars will be significantly enhanced due to three-body scattering \citep{Chen_2009}, which often places the captured stars on highly eccentric orbits around the SMBH.

\begin{figure}[h]
\centering
\includegraphics[width=0.9\linewidth]{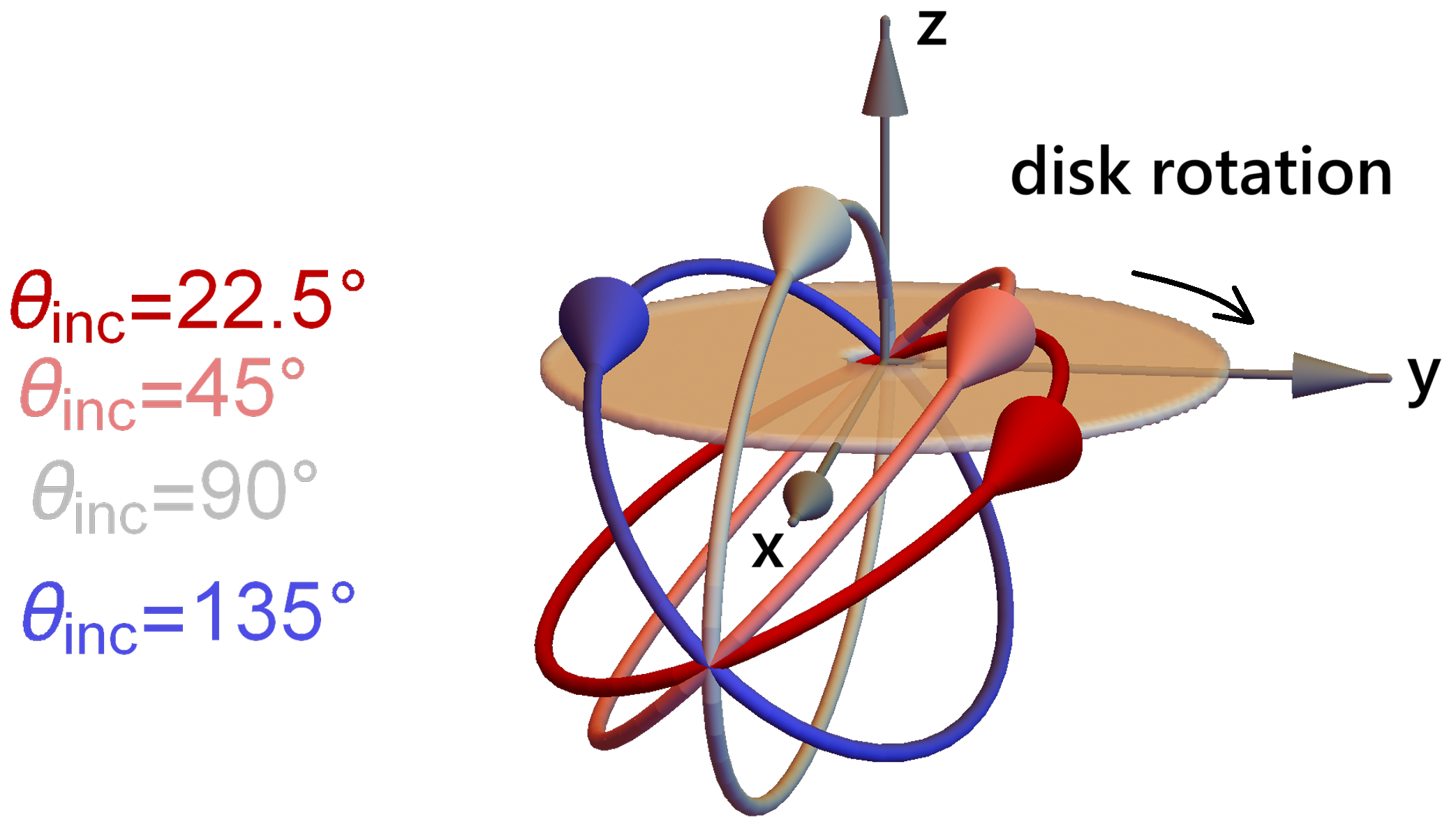}
\caption{
\label{orbital_inc} 
A sketch map illustrating the inclined stellar orbits adopted in this work. Each orbit is color-coded according to its inclination angle, with an arrow indicating the orientation of motion. The direction of rotation of the accretion disk is marked with a black arrow for reference.
}
\end{figure}
To investigate the dependence on orbital inclination, we fix the semi-major axis of the stellar orbit along the $x$-axis and rotate the orbital plane to inclination angles of $\theta_{\rm inc} = 22.5^\circ$, $45^\circ$, $90^\circ$, and $135^\circ$ relative to the $x$-$y$ plane (from models R22.5 to R135), as illustrated in Figure \ref{orbital_inc}.
In addition to previous studies that focused on either in-disk \citep{2024MNRAS.527.8103R} or perpendicular \citep{2019ApJ...881..113C} encounters, our setup includes various orbital inclinations, providing a more general scenario of TDE-disk interactions.

Among these four models, the projection of the stellar orbital angular momentum onto the disk plane varies from prograde to retrograde relative to the disk angular momentum. 
Specifically, in the $22.5^{\circ}$ and $45^{\circ}$ cases (hereafter projected prograde, where the projected stellar orbit is aligned with the disk rotation), the stellar angular momentum has a component aligned with that of the disk. 
At $90^{\circ}$, the two are perpendicular. 
In the $135^{\circ}$ case (hereafter projected retrograde, where the projected orbit is anti-aligned with the disk), the stellar angular momentum has a component opposite to that of the disk.

Additionally, for the two cases of $\theta_{\rm inc}=45^\circ$ and $90^\circ$, we apply both Newtonian potential (model N45 and N90) and GR potential (model R45 and R90) to examine the impact of apsidal precession.
To ensure a reliable comparison, the initial conditions for both the Newtonian and GR models are kept identical. Detailed information of our parameter space is summarized in Table \ref{table:setups}.

\section{Hydrodynamical Results}\label{sec:numerical_results}
\subsection{Overview}

\begin{figure*}[ht]
\centering
\includegraphics[width=0.95\linewidth]{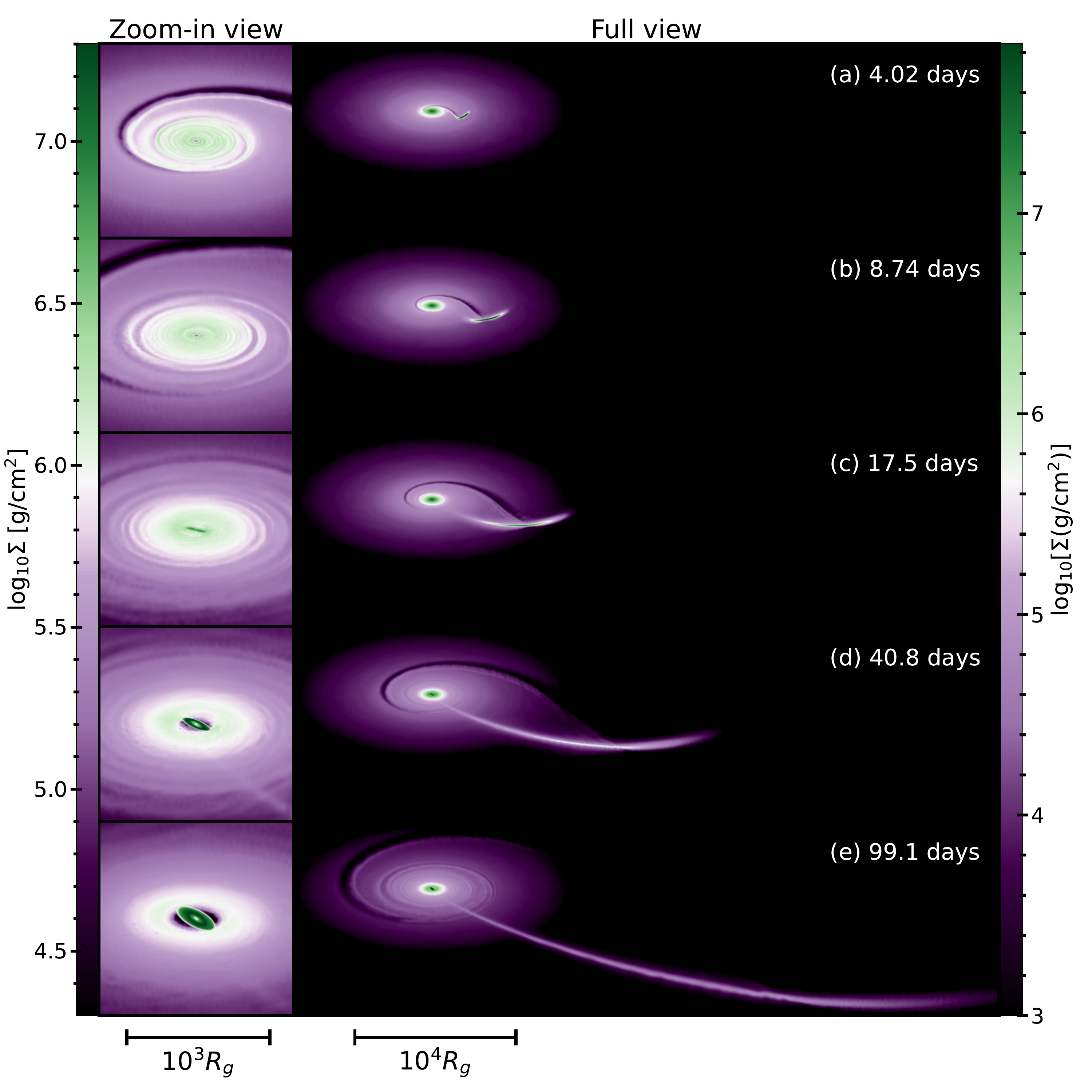}
\caption{\label{overview} 
Time evolution of the log-scaled column density in an arbitrary inclined view of our fiducial model R90 ($\theta_{\rm inc}=90^\circ$) over $\sim$100 days. 
Each snapshot presents a zoom-in view of the inner region on the left and a full view of the co-evolution between the outer disk and the stellar debris on the right.
Time since the start of the simulation is labeled in each snapshot. 
Spatial scale bars for the zoomed-in and full views are displayed beneath the figure.
(a)-(b): the star is disrupted and tears through the disk, leaving a gap on the outer disk along its trajectory.
(c): The debris separates from the disk, with its leading end beginning to fall back toward the pericenter. This marks the end of continuous debris-disk interaction and the onset of the fallback phase.
(d)-(e): The fallback stream rapidly reaches its peak intensity and subsequently weakens as the debris continues to be stretched by tidal forces. An inclined inner disk forms during the fallback phase.
An animated version of this figure is available \href{https://www.youtube.com/watch?v=hIt90J9X8uY}{here}, showing the face-on column density from 0 to $\sim100$ days across four inclination angles ($\theta_{\rm inc}=22.5^\circ,\ 45^\circ,\ 90^\circ$, and $135^\circ$).
}
\end{figure*}
We present the co-evolution of the debris and the AGN disk lasting for $\sim100$ days of our fiducial model (R90) in Figure \ref{overview}. 

During the disruption phase,
as the debris increases in size and decreases in density, it tears through the outer disk,   clearing out a region and becoming diffused while leaving the black hole, similar to the in-plane TDE scenario in \cite{2024MNRAS.527.8103R}. 
This perturbation from the first pericenter passage to the debris's emergence from the AGN disk may produce a precursor burst before the TDE flare, as also predicted by \cite{2024MNRAS.527.8103R}. 
This may serve as a distinctive signature of AGN TDEs and will be further discussed in Section \ref{sec:precursor}.
Similar to the in-disk scenario in \cite{2024MNRAS.527.8103R}, the structure of the debris in our fiducial model deviates from that of a naked TDE due to continuous interaction with the disk.
However, this ongoing interaction after the first pericenter passage neither causes complete mixing of the debris into the disk like some cases in \cite{2024MNRAS.527.8103R}, nor does it cause severe disruption of the outer disk, since only the portion of the stream interacting with the disk is affected.

During the fallback phase, our fiducial model does not exhibit complete destruction of the inner region of the disk as reported in \cite{2019ApJ...881..113C}, although the inflow rate is enhanced due to perturbations (see Section \ref{sec:inflow} for details). 
Instead, a misaligned new inner disk forms, as shown in the left column of Figure \ref{overview}.

We attribute this difference to the following possible factors.
First and foremost, the intensity of the debris is a controlling parameter in \cite{2019ApJ...881..113C}, which remains fixed during the stream collision.
However, the fallback stream in our simulation quickly reaches its peak flux and then declines to a relatively low level for most of the fallback phase, as the debris continues to be stretched by tidal forces (see the right column of Figure \ref{overview}). 
As a result, the returning stream remains too weak to effectively dissipate the angular momentum of the inner disk, leading to gradual accumulation and the formation of a misaligned inner disk.
Second, the aspect ratio of the disk $H/R\approx0.01$ in our simulation is significantly smaller than $H/R\approx0.1$ as in \cite{2019ApJ...881..113C}, yielding a much denser inner region that is more resistant to perturbations.
Third, the efficient radiative cooling term prevents the sudden rise in internal energy due to stream collisions from feeding back into the disk through $PdV$ work, thereby reducing its dynamical impact on the disk structure.
Finally, the debris in the parabolic orbits used in \cite{2019ApJ...881..113C} carries more orbital energy than in the elliptical cases considered here. 

\subsection{The Dependence on Orbital Inclination}\label{subsec:Inc}
In this section, we discuss the effect of the star's orbital inclination on stream collisions in AGN TDEs.
Since the stellar orbital inclination determines whether the projection of the angular momentum of the fallback material onto the disk plane aligns, is perpendicular to, or opposes the angular momentum of the disk, we expect the outcome of the collision to exhibit a significant dependence on the orbital inclination.

\subsubsection{Morphology}
\begin{figure*}[ht]
\centering
\includegraphics[width=0.9\linewidth]{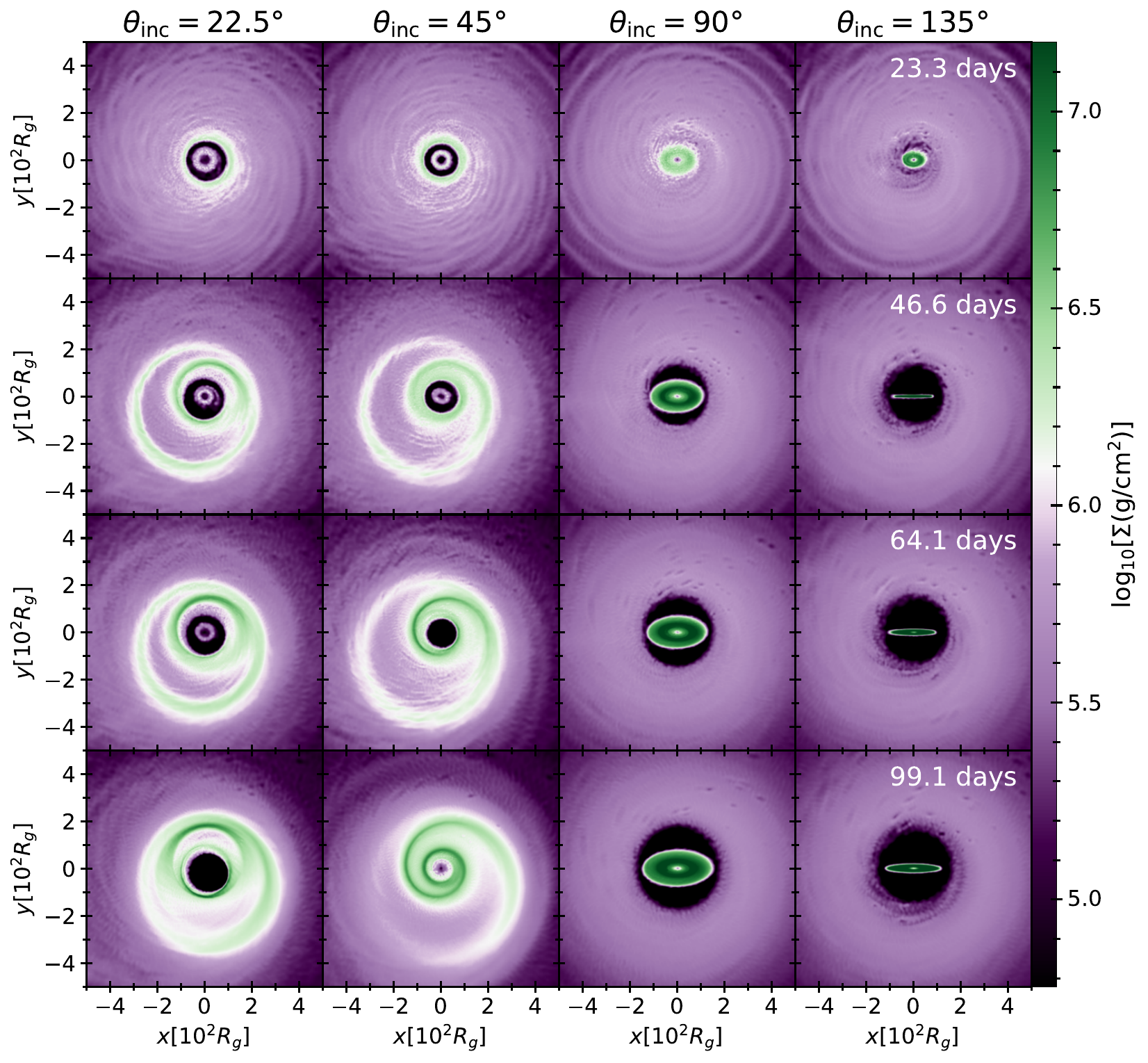}
\caption{\label{innerdisk}
Face-on views of the evolution of the log-scaled column density of the inner region ($R < 10^3 R_g$) of the disk for models with $\theta_{\rm inc} = 22.5^\circ$, $45^\circ$, $90^\circ$, and $135^\circ$. 
The $22.5^\circ$ and $45^\circ$ models develop a central cavity and prominent spiral arms, while the $90^\circ$ and $135^\circ$ models form a gradually tilting inner disk.
The time since the start of the simulation for each snapshot is labeled on the top right of each row. 
}
\end{figure*}

As the orbital inclination ($\theta_{\rm inc}$) shifts from projected prograde ($22.5^\circ$, $45^\circ$) to perpendicular ($90^\circ$) and then to projected retrograde ($135^\circ$), the interaction between the fallback debris and the AGN disk gives rise to two distinct inner disk configurations: a central cavity with strong spirals around it, or a misaligned inner disk, as shown in Figure \ref{innerdisk}.

For $\theta_{\rm inc}=22.5^\circ$ and $45^\circ$, the fallback of the debris stream clears out the disk inside $R_p\approx70R_g$ while inducing spiral shock outside the cavity within $\sim500R_g$ from the SMBH. 
The spiral is stronger and the cavity is larger in the $22.5^\circ$ case than in the $45^\circ$ case.
As the stream weakens in the later stages of fallback, the material outside the impact radius flows inward with oscillation due to apsidal precession and rapidly refills the cavity in the $45^{\circ}$ case.

For $\theta_{\rm inc}=90^{\circ}$ and $135^{\circ}$, the material from both the disk and the fallback debris gathers inward due to the loss of angular momentum during the impact process. 
As time goes on, an inclined inner disk forms relative to the original AGN disk. 
As the debris continues to fall back, the inner disk grows in size with a gradually increasing inclination angle.

\subsubsection{Angular Momentum Transfer, Inflow Rate, and Radial Profile Evolution}\label{sec:inflow}
\begin{figure*}[ht]
\centering
\includegraphics[width=0.88\linewidth]{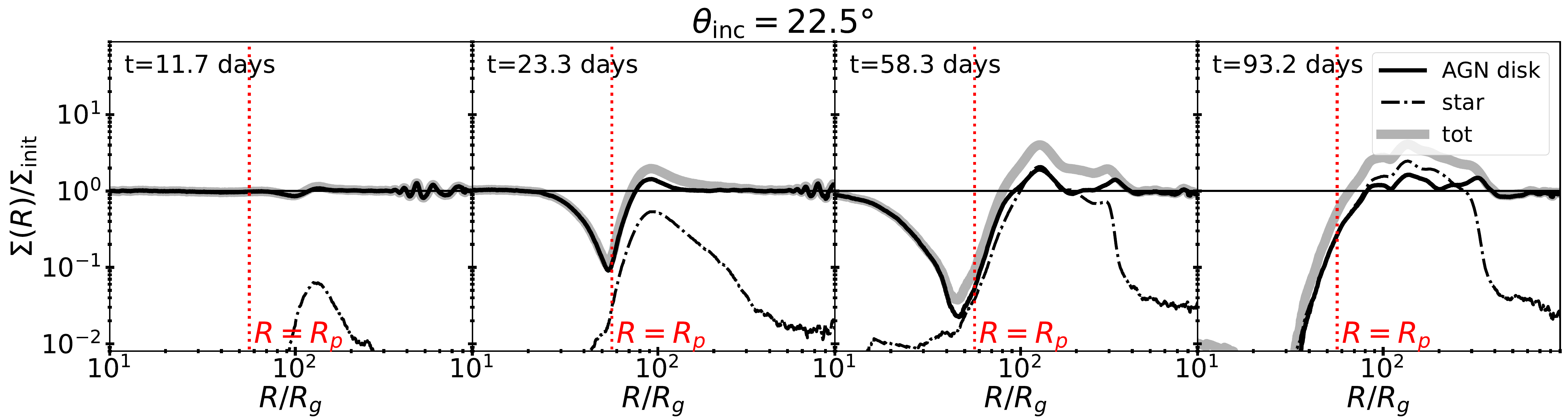}
\includegraphics[width=0.88\linewidth]{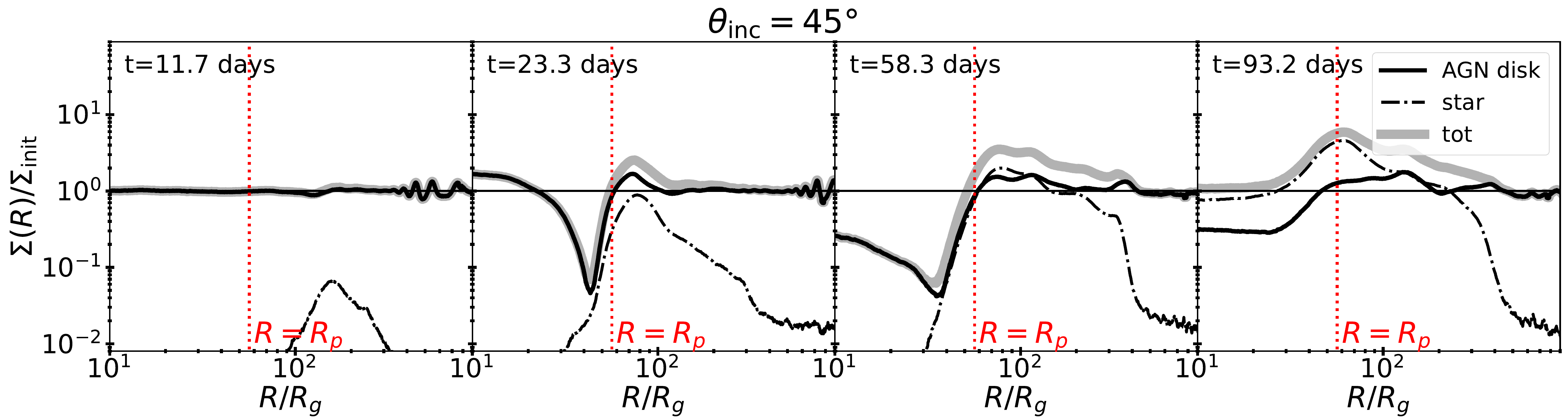}
\includegraphics[width=0.88\linewidth]{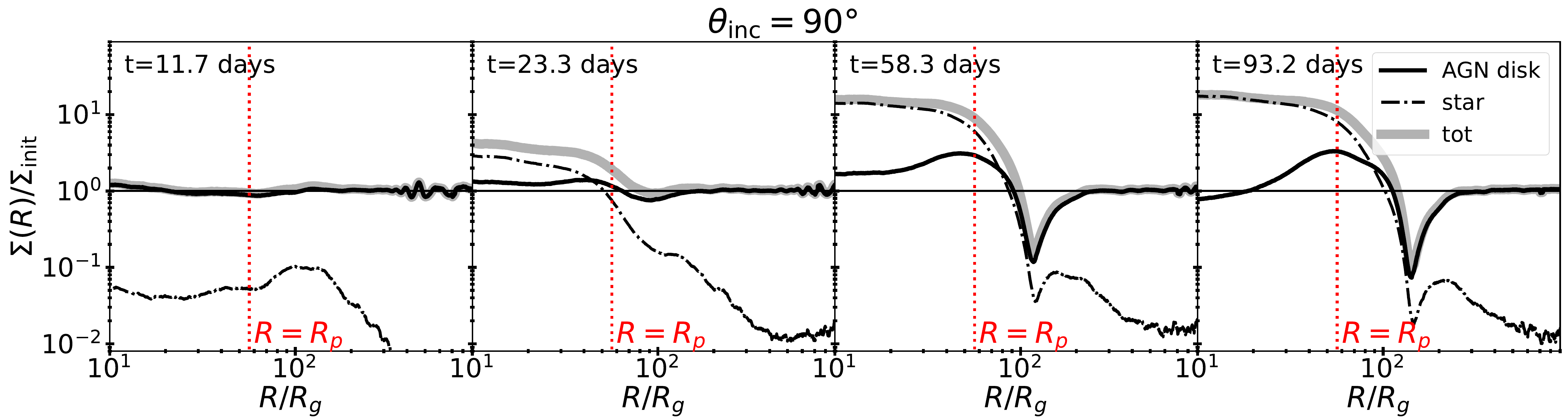}
\includegraphics[width=0.88\linewidth]{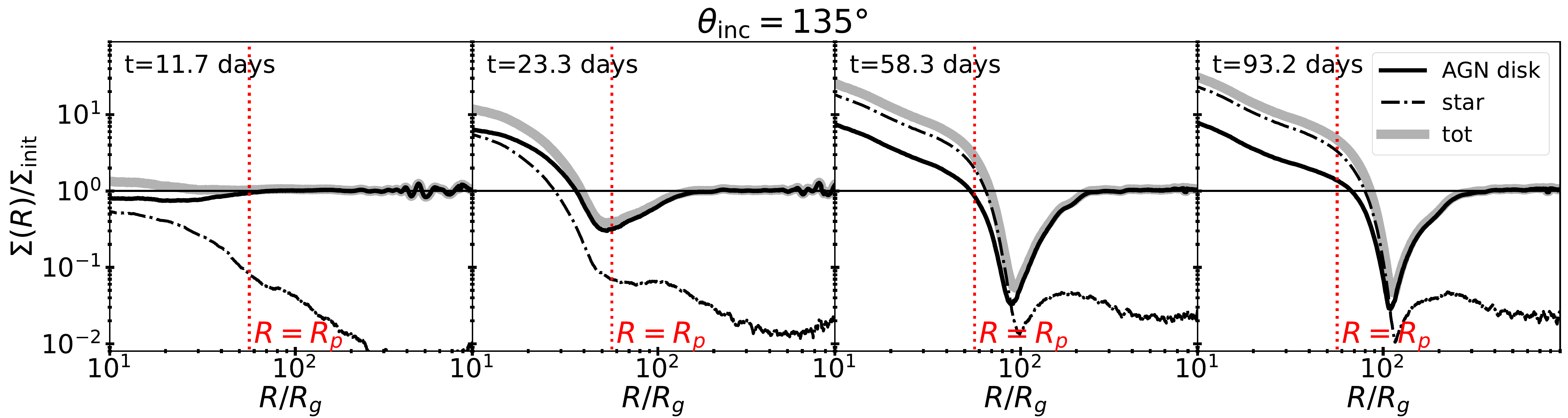}
\caption{\label{dens_profile} 
Time evolution of the radial surface density profiles for models with stellar orbital inclinations of $22.5^\circ$, $45^\circ$, $90^\circ$, and $135^\circ$. The plot range is limited to the inner region ($R<10^3R_g$). The total surface density, AGN disk contribution, and stellar debris contribution are shown by the solid gray, solid black, and dash-dotted black lines, respectively. The pericenter distance as well as the impact radius ($R_p$) is indicated by dashed red vertical lines. 
The corresponding time for each panel is labeled in the upper-left corner.
}
\end{figure*}
\begin{figure*}[ht]
\centering
\includegraphics[width=\linewidth]{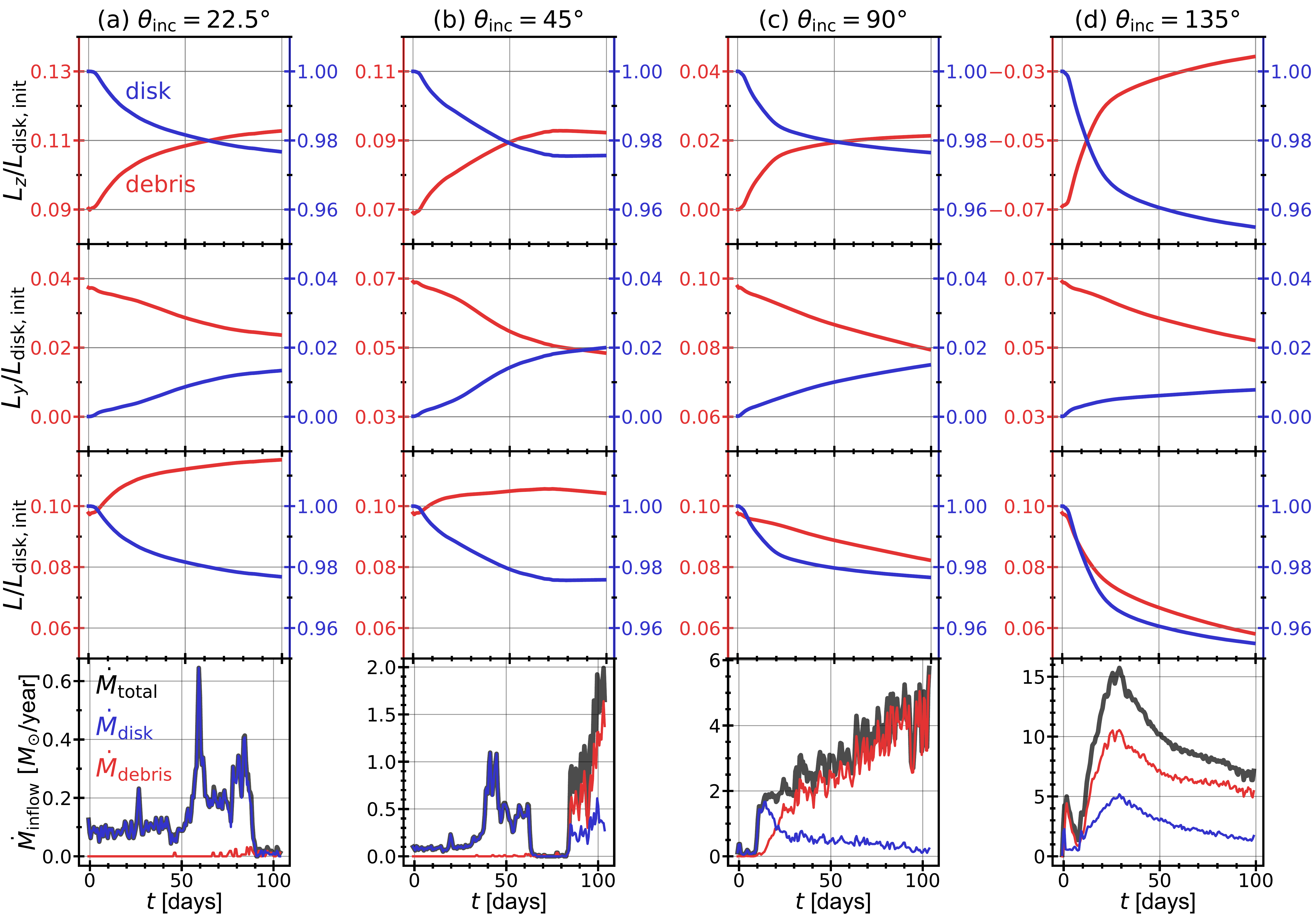}
\caption{\label{AM_and_inflow_rate} 
Time evolution of angular momentum and inflow rate for the stellar debris and the AGN disk, for models with orbital inclinations $\theta_{\rm inc} = 22.5^\circ$, $45^\circ$, $90^\circ$, and $135^\circ$ from (a) to (d). 
The top three rows show the angular momentum, normalized by the disk's initial angular momentum ($L_{\rm disk,init}$). 
From top to bottom are the $z$-component $(L_z)$, $y$-component $(L_y)$, and $L=\sqrt{L_x^2+L_y^2+L_z^2}$ with $L_x$ denoting the $x$-component. 
Angular-momentum components of the stellar debris (red, left axis) and the disk (blue, right axis) are shown, with a dynamically adjusted and a fixed plotting range, respectively.
The bottom row shows the inflow rate as a function of time, with the total inflow rate ($\dot{M}_{\rm total}$), stellar debris contribution ($\dot{M}_{\rm debris}$), and disk contribution ($\dot{M}_{\rm disk}$) plotted in black, red, and blue, respectively.
}
\end{figure*}

The progression of azimuthally averaged surface density $\Sigma(R)$ over time in Figure \ref{dens_profile}, along with the temporal evolution of angular momentum and inflow rate in Figure \ref{AM_and_inflow_rate}, help us understand the formation of different inner disk structures due to various $\theta_{\rm inc}$ as presented in Figure \ref{innerdisk}.

The angular momentum components in $z$-direction ($L_z$, parallel to the AGN disk's orientation) 
and $y$-direction ($L_y$, aligned with the stellar orbital orientation of the $\theta_{\rm inc}=90^\circ$ case), 
along with the total angular momentum $L=\sqrt{L_x^2+L_y^2+L_z^2}$ and the inflow rate $\dot{M}_{\rm inflow}$ of both the debris and the disk components, are shown in Figure \ref{AM_and_inflow_rate}, from top to bottom.
For all four $\theta_{\rm inc}$, the stellar debris gains $L_z$ from the disk, 
while $L_y$ is transferred from the stellar debris to the disk due to the inclination of the orbit.
However, the evolution of $L$ (the third row of Figure \ref{AM_and_inflow_rate}) of the stellar debris and the disk varies across different cases.

In the projected prograde cases ($22.5^\circ$ and $45^\circ$), the debris extracts angular momentum from the disk upon stream-disk collision, leading to an increase in $L_{\rm debris}$.
Consequently, the returning stellar debris accumulates outside $R_p$ (see Figures \ref{dens_profile}(a) and (b)), while the disk loses angular momentum and the inflow is intensified. 
The sustained fallback further creates a gap around $R_p$, hindering the accretion supply and rapidly clearing out the area within $R_p$.
These altogether cause the inflow to be dominated by disk material (see Figures \ref{AM_and_inflow_rate}(a) and (b)).
The inflow temporarily ceases as the inner region is completely emptied.
​Only after this cavity is refilled does the stellar debris begin to dominate the inflow.

Conversely, in the perpendicular ($\theta_{\rm inc}=90^{\circ}$) and projected retrograde ($\theta_{\rm inc}=135^{\circ}$) cases, counter-streaming collisions lead to more violent angular momentum exchange while simultaneously reducing the angular momentum of both the stellar debris ($L_{\rm debris}$) and the disk ($L_{\rm disk}$) (see the third row of Figure \ref{AM_and_inflow_rate} (c) and (d)). 
This significantly enhances the inflow, consistent with the phenomenon shown in Figures \ref{dens_profile} (c) and (d) that both stellar debris and disk material accumulate within $R_p$ during the fallback phase. 
As fallback goes on, the inner disk becomes increasingly dominated by stellar debris while expanding outward. 
In these cases, the inflow is primarily composed of stellar debris and reaches roughly an order of magnitude higher than in the projected prograde cases.

\subsection{The Role of Relativistic Apsidal Precession}\label{sec:GR}
For SMBHs with mass $M_{\rm BH}\approx10^6M_\odot$, the tidal radius for a main-sequence (MS) star is only a few tens to hundreds of gravitational radii $(R_g=GM_{\rm BH}/c^2)$, implying that the star must pass very close to the black hole to be disrupted. 
In this regime, GR effects on stellar orbits become important.  
While the orbital deviation of the debris center of mass in the Schwarzschild metric relative to Newtonian gravity may be moderate, the expansion of the debris after disruption amplifies differences in the trajectories due to relativistic precession, potentially leading to markedly different outcomes in stream collisions.

To investigate how orbital precession affects debris-disk interactions, we compare the outcomes of two pairs of our models with different orbital inclinations, each simulated under both the Newtonian and GR potentials (model R45 vs. N45, model R90 vs. N90).
These particular cases serve as illustrative examples to demonstrate that, in the AGN TDE scenarios, stream collisions are influenced by GR-induced apsidal precession in a more complex fashion than in naked TDEs, underscoring the importance of including relativistic effects.

\subsubsection{Debris-disk collision during the disruption phase}
\begin{figure*}[ht]
\centering
\includegraphics[width=0.71\linewidth]{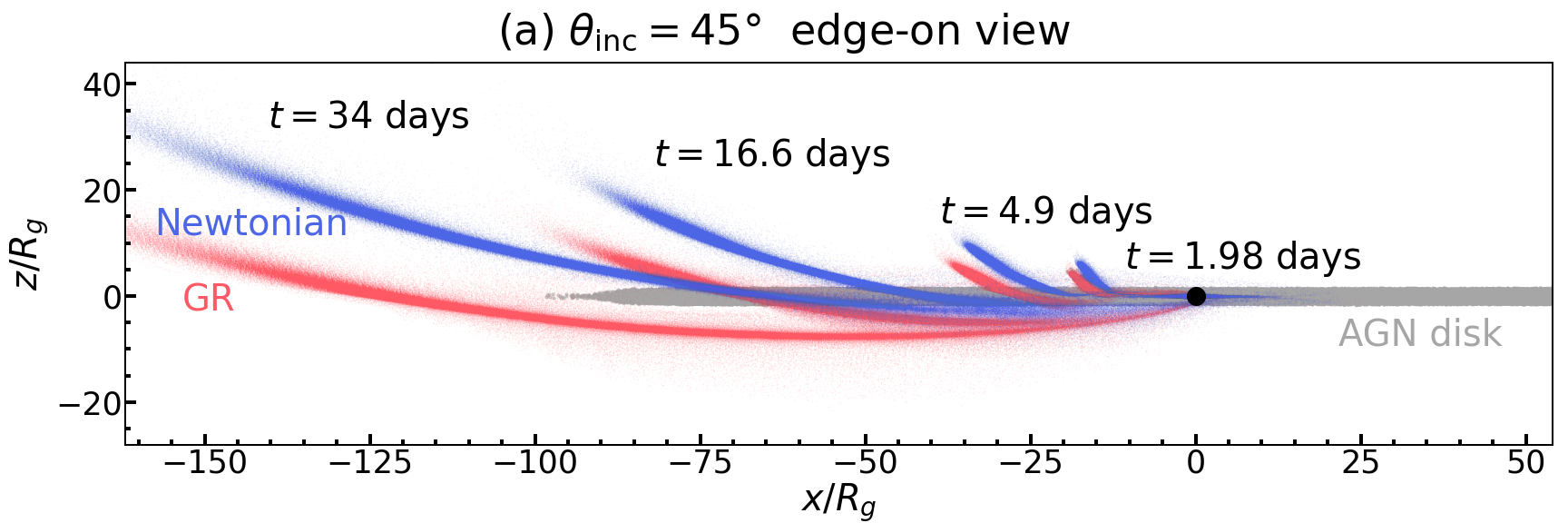}
\includegraphics[width=0.71\linewidth]{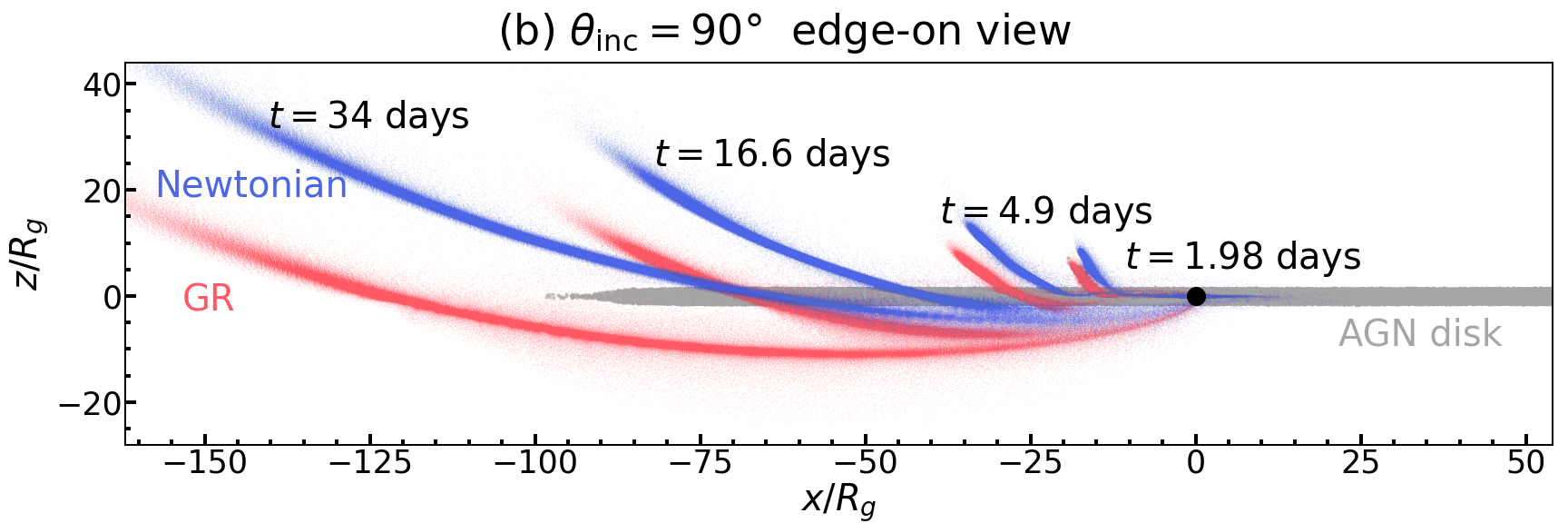}
\caption{
\label{GR1}
Edge-on views illustrating the debris under Newtonian gravity (blue) and Schwarzschild metric (red) at $t=1.98$ days, $4.9$ days, $16.6$ days and $34$ days.
Panels (a) and (b) correspond to stellar orbital inclinations $\theta_{\rm inc} = 45^\circ$ and $\theta_{\rm inc} = 90^\circ$, respectively, relative to the AGN disk plane. 
In the Schwarzschild metric, relativistic apsidal precession causes the debris to leave the disk earlier, resulting in a shorter duration of debris-disk interaction than Newtonian cases. 
}
\end{figure*}
During the disruption phase, the debris-disk interaction differs significantly between the Newtonian and GR cases, resulting in distinct morphologies for both the outer disk and the stellar debris.

\begin{figure*}[ht]
\centering
\includegraphics[width=0.95\linewidth]{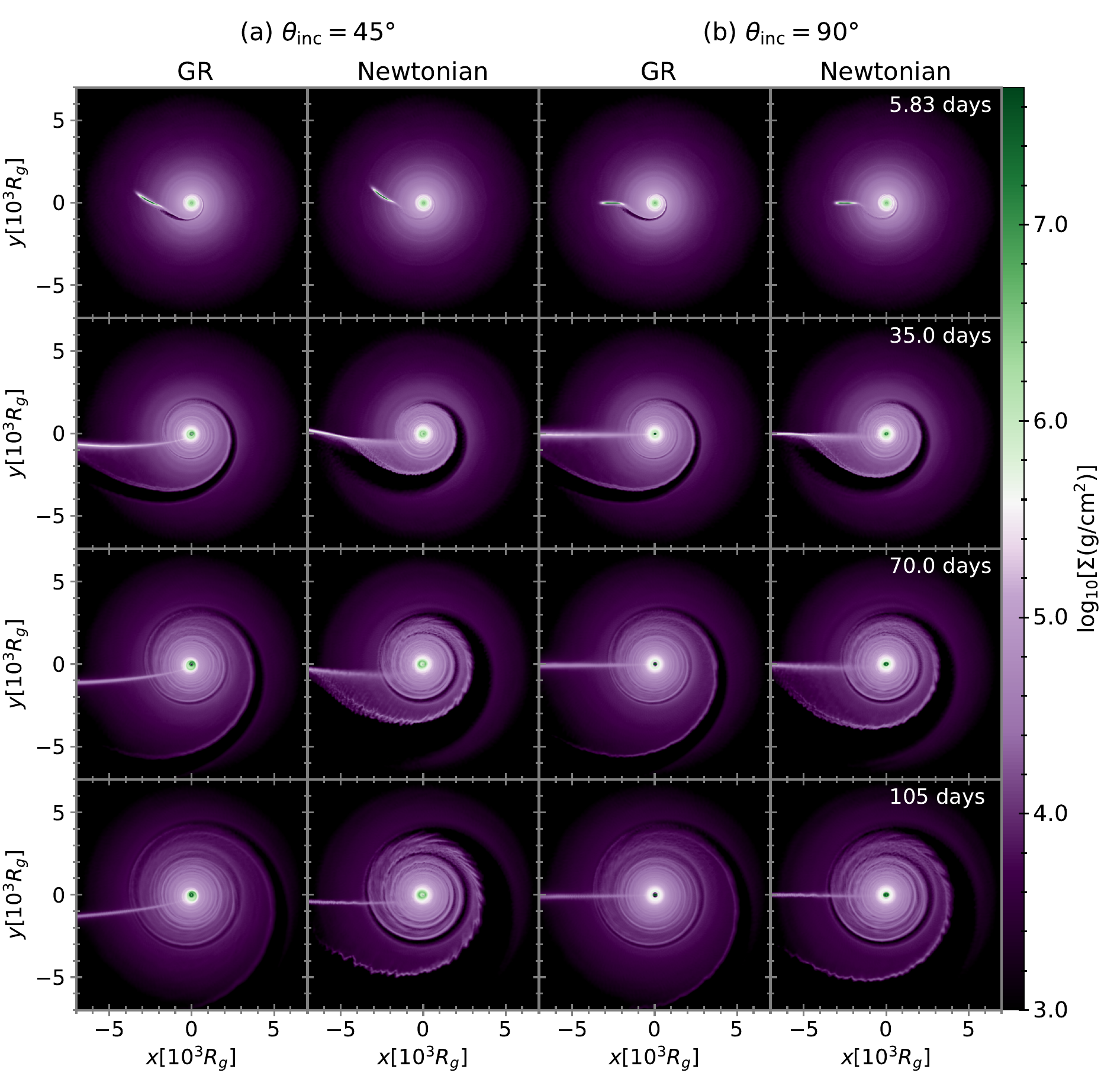}
\caption{
\label{GR2}
Face-on views illustrating the different outer disk structures developed under GR and Newtonian gravity at $t=5.83$ days, $35$ days, and $105$ days, from top to bottom.
Panels (a) and (b) correspond to stellar orbital inclinations $\theta_{\rm inc} = 45^\circ$ and $\theta_{\rm inc} = 90^\circ$, respectively, relative to the AGN disk plane. 
In the Schwarzschild metric, a less turbulent outer disk compared to the Newtonian cases is developed. 
An animated version of this figure is available \href{https://youtube.com/shorts/5UlTGsKn2E0}{here}, displaying the column density evolution under GR and Newtonian gravity from 0 to $\sim106$ days. 
Both face-on and edge-on views are shown, with the left two columns providing zoom-in views of the inner disk structure corresponding to the right two columns.
}
\end{figure*}

Specifically, in the Schwarzschild metric, relativistic apsidal precession modifies the orbital trajectory of the debris, causing it to exit the disk region earlier than in the Newtonian case, as illustrated in Figure \ref{GR1}.
This difference in trajectory shortens the duration of the debris-disk collision in the GR case
and further leads to distinct outcomes, as shown in Figure \ref{GR2}. 
In the GR case (right columns of Figure \ref{GR2}(a) and (b)), due to early termination of debris-disk interaction, the outer disk remains comparatively  less turbulent, while the debris stream appears more coherent.
In contrast, the Newtonian case (left columns of Figure \ref{GR2}(a) and (b)) features a more prolonged and violent interaction, resulting in a more strongly perturbed outer disk and a fragmented debris stream with clumpy substructures.
Additionally, the fallback material also returns at a noticeably different angle due to precession in the GR case. 

\subsubsection{Inner disk evolution during fall-back phase}
\begin{figure*}[ht]
\centering
\includegraphics[width=0.95\linewidth]{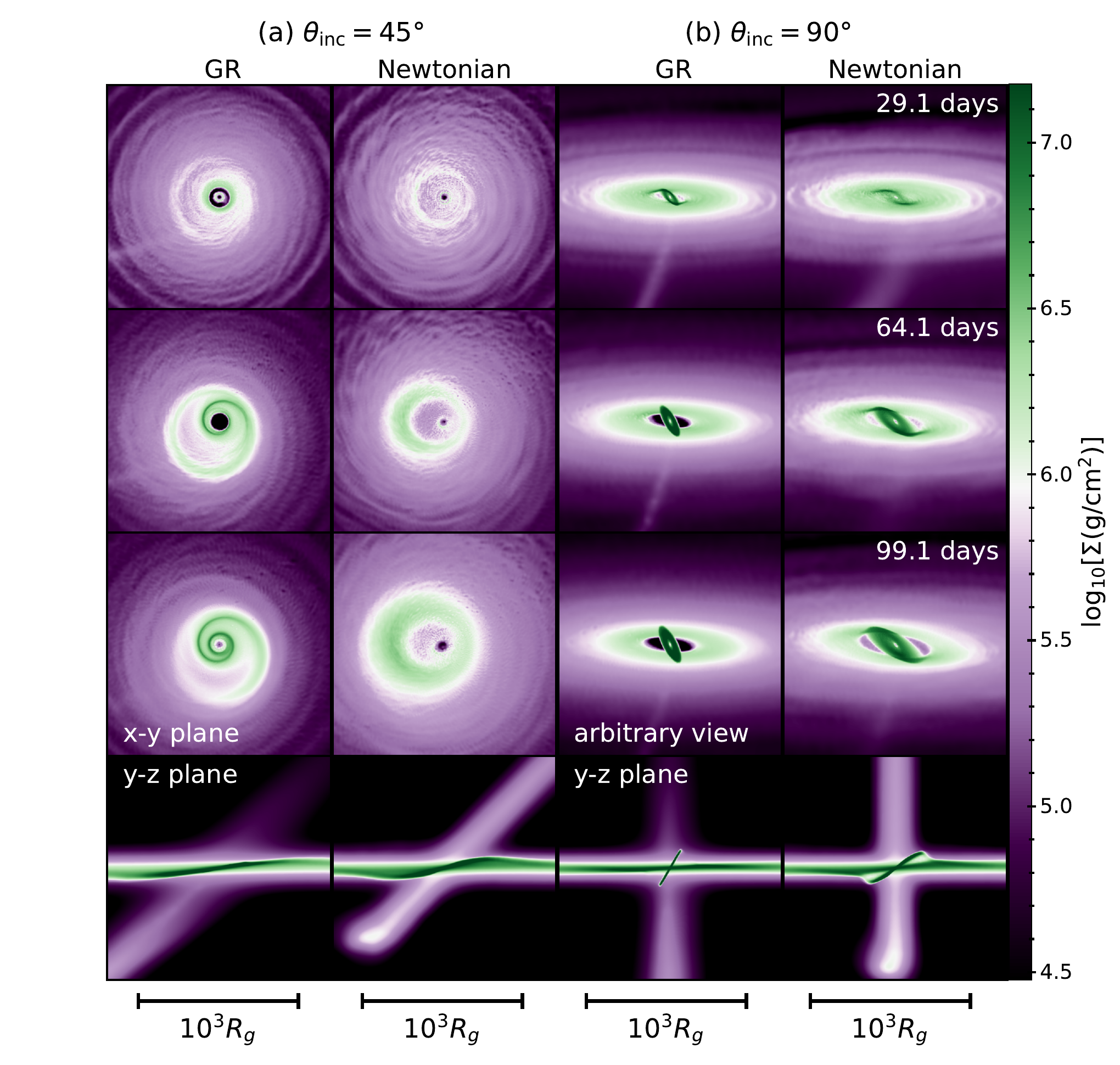}\hspace{-1.2mm}
\caption{\label{Sch_Kep_inner_disk} 
Comparison of the column density distributions of the inner disk under Newtonian gravity (left panels) and Schwarzschild metric (right panels), color-coded on a logarithmic scale. 
Panel (a) shows the case of $\theta_{\rm inc} = 45^\circ$, viewed face-on relative to the initial AGN disk plane, and panel (b) shows the case of 
$\theta_{\rm inc} = 90^\circ$, viewed from an arbitrary inclined angle to better reveal the formation of a tilted inner disk. 
For each case, the bottom row displays the projected column density in the $y$-$z$ plane, providing an edge-on view of the inner disk structure.}
The time since the start of the simulation is marked in each panel, and a scale bar of $10^3R_g$ is shown for reference.
\end{figure*}
In addition to differences in the early evolution of the debris and outer disk structure, the GR and Newtonian potentials also lead to distinct inner disk configurations during the fallback phase.

Specifically, for $\theta_{\rm inc} = 45^\circ$ (Figure \ref{Sch_Kep_inner_disk}(a)), spiral shocks develop around $\sim300R_g$ in the GR case.
The region within $R_p$ is cleared due to perturbation and a lack of accreting supply, and is subsequently refilled by unstable inflow from fluid undergoing relativistic precession.
In contrast, in the Newtonian case, relatively inefficient angular momentum dissipation prevents the fallback material from inflowing, leading to its accumulation in an elliptical region near $\sim300R_g$, while the inner region within $R_p$ remains filled.
For $\theta_{\rm inc}=90^\circ$ (Figure \ref{Sch_Kep_inner_disk}(b)), the inner disk formed in the Schwarzschild metric is more compact and denser, with a noticeably larger inclination angle compared to that under Newtonian gravity.

We attribute these differences to several key factors.
Most importantly, the deviation between the GR and Newtonian gravitational potentials becomes increasingly significant closer to the black hole.
In the GR framework, non-circular trajectories are subject to relativistic apsidal precession. 
Perturbations due to fallback stream can excite orbital eccentricity of the fluid, which in turn lead to hydrodynamic instabilities and obital collisions that greatly enhance angular momentum transport \citep{Deng_2025}.
Additionally, subtle differences in debris trajectories result in a shorter debris-disk interaction phase in the GR case, ultimately producing a more coherent fallback stream.
This, combined with a slightly altered injection direction caused by apsidal precession, may also contribute to the observed discrepancies in the inner disk structure.

\section{Synthetic Multi-band Variability}\label{sec:Synthetic}
In this section, we present the radiative signatures derived from the hydrodynamical simulation results, including the characteristics of the light curves and multi-band features produced by blackbody emission from the accretion disk.

\subsection{Radiative Post-processing Method}
To construct synthetic multi-band light curves, we assume that a blackbody spectrum is emitted from the disk, with each annulus radiating at a fixed effective temperature equal to the blackbody effective temperature ($T_{\rm eff}(R)$) of the unperturbed AGN disk.
To estimate the time-dependent radiative power, we note that most of the internal energy generated from the dissipation of orbital energy is promptly removed from the system via the implemented radiative cooling term. Therefore, after neglecting the effect of advection, the total energy loss rate from the system provides a reasonable approximation for the emitted luminosity.
Specifically, we compute the local Lagrangian energy loss by tracking the same group of fluid elements over time to obtain the radial profile of energy dissipation rate, denoted as $dE_{\rm{tot}}(R)/(dtdS)$--the total energy loss per unit area per unit time at radius $R$. 
For each annulus, the radiative power is then assumed to be the local energy dissipation rate.
A detailed description of the calculation of the synthetic variability can be found in Section \ref{sec:energy_loss_calc}.

\subsection{The two-phase light curve}\label{sec:variability}
\begin{figure*}[ht]
\centering
\includegraphics[width=\linewidth]{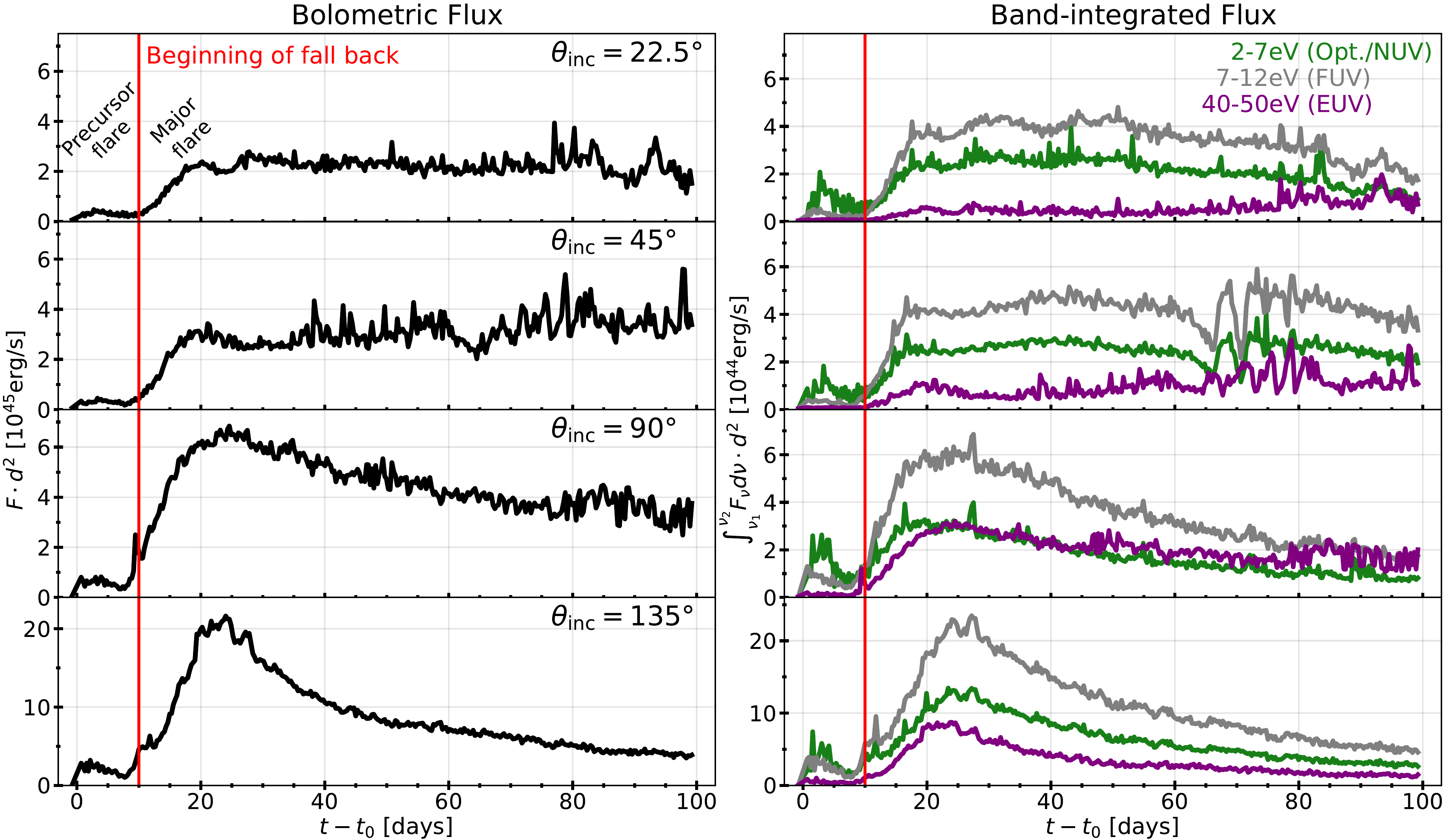}
\caption{\label{precursor} 
Synthetic light curves exhibiting a two-phase structure: a precursor flare, associated with the process of debris tearing through the AGN accretion disk, followed by a major flare powered by the fall-back of debris.
Time is measured from the first pericenter passage ($t_0 \approx 0.73$ days).
The onset of the fall-back phase, marking the transition between the two flares, is indicated by a red vertical line in each panel.
All panels present a face-on view of the system. 
Each row corresponds to a different orbital inclination angle, $\theta_{\rm inc} = 22.5^\circ$, $45^\circ$, $90^\circ$, and $135^\circ$, from top to bottom.
The left column displays the bolometric luminosity ($F \cdot d^2$, where $F$ is the flux at an arbitrary reference distance $d$), while the right column shows its band-integrated components in three energy ranges: 2-7 eV (green), 7-12 eV (gray), and 40-50 eV (purple), corresponding to the NUV/optical, FUV, and EUV bands, respectively.
}
\end{figure*}
In the synthetic light curves produced by our simulations (Figure \ref{precursor}), a distinct two-phase structure emerges, comprising a precursor flare followed by a major flare.
Overall, the former has a peak flux several times lower and a duration several times shorter than that of the latter.
In terms of morphology, the precursor flare exhibits a more symmetric profile, while the major flare is characterized by a TDE-like rapid rise and a gradual decline.

\subsubsection{Early debris-disk interaction: the precursor flare}\label{sec:precursor}
The precursor flare is temporally consistent with the process when the disrupted star leaves the SMBH while tearing through the disk, corresponding to the dynamical stages shown in Figure \ref{overview}(a)-(c). 

As this continuous debris-disk interaction mainly disturbs the outer regions of the disk, the precursor flare is prominent in the optical/UV bands but diminishes in the EUV band. 
As shown in Figure \ref{precursor}, the bolometric flare intensity increases with orbital inclination, from projected prograde ($\theta_{\rm inc} = 22.5^\circ$) to projected retrograde ($\theta_{\rm inc} = 135^\circ$), and the corresponding spectral peak exhibits a shift toward higher energies at higher inclinations.
This is expected as shocks become more prominent when the orbital velocity difference increases.

\subsubsection{Fall-back phase: the major flare}\label{sec:major_flare}
The major flare emerges at the onset of the fallback process, corresponding to the stages depicted in Figure \ref{overview}(c)-(e). 
It exhibits a typical TDE-like light curve characterized by a rapid rise and a slower decay.

As shown in the left panels in Figure \ref{precursor}, the bolometric luminosities estimated from the total energy loss rate $(\dot{E}_{\rm tot})$ vary noticeably across different inclination angles. 
Specifically, both the projected retrograde case ($\theta_{\rm inc}=135^\circ$) and the perpendicular case ($\theta_{\rm inc}=90^\circ$) exhibit much higher peak luminosity, which are several times greater than those in the projected prograde cases ($\theta_{\rm inc}=22.5^\circ$ and $45^\circ$). 
This may be due to more violent counter-streaming collisions that result in stronger orbital energy dissipation. 
However, these flares corresponding to perpendicular and projected retrograde orbits are also much shorter in duration compared to the projected prograde cases, which exhibit longer-lasting flares with dimmer peak luminosity. 
This trend is consistent with the prediction by \citet{McKernan_2022}, suggesting that retrograde TDEs in AGNs produce brighter but shorter-lived flares than prograde ones.
The spectral peak lies in the UV band, slightly harder than that of the precursor flare. The multi-wavelength variability characteristics of the major flare will be discussed in detail in Section \ref{sec:innerdisk_lc}.

\subsection{Multi-band signatures}\label{sec:innerdisk_lc}
\begin{figure*}[ht]
\centering
\includegraphics[width=\linewidth]{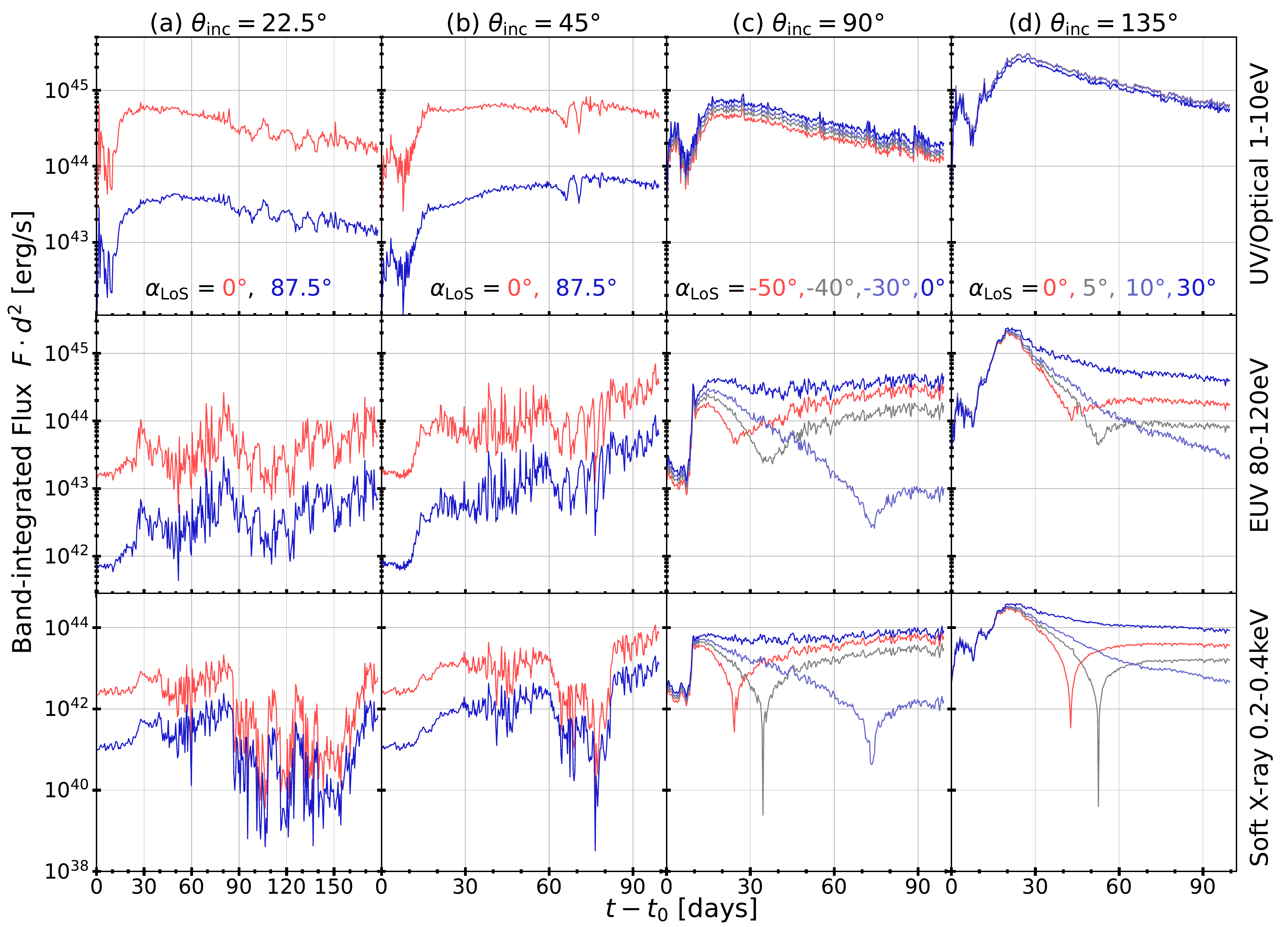}
\caption{\label{inner} 
Comparison of the synthetic multi-band light curves ($F \cdot d^2$, with $F$ being the flux at reference distance $d$) for different orbital inclinations ($\theta_{\rm inc}$) and viewing angles relative to the initial AGN disk plane ($\alpha_{\rm LoS}$). 
Time is measured from the first pericenter passage ($t_0 \approx 0.73$ days).
Each column corresponds to an orbital inclination of $\theta_{\rm inc} = 22.5^\circ$, $45^\circ$, $90^\circ$, and $135^\circ$, from left to right. 
Each row is for a certain energy band: 1-10 eV (top), 80-120 eV (middle), and 0.2-0.4 keV (bottom), corresponding to the UV/optical, EUV, and soft X-ray bands, respectively. 
In each panel, different viewing angles ($\alpha_{\rm LoS}$) are color-coded, as indicated in the top panel of each column.
}
\end{figure*}
In this section, we discuss the synthetic multi-band signatures derived from our simulation data which implies several distinctive observational phenomena.
Assumptions and detailed procedure for computing the time-dependent disk continuum are in Section \ref{sec:spec_calc}.
The time evolution of the band-integrated flux variations in three energy bands (1-10eV, 80-120eV, and 0.2-0.4keV) are presented in Figure \ref{inner} for different orbital inclinations $(\theta_{\rm inc})$ and viewing angles $(\alpha_{\rm LoS})$ relative to the initial AGN disk.
Beyond reproducing an expected TDE-like UV/optical flare with a decline rate sensitive to the orbital inclination $\theta_{\rm inc}$, our synthetic spectrum predicts distinct observational signatures in the evolution of disk blackbody component in EUV/soft X-ray band among various inner disk configurations across different $\theta_{\rm inc}$. 

In the inner disk depletion scenario (the projected prograde cases, models R22.5 and R45), UV and soft X-ray emission from the innermost region within $\sim R_p$ is temporarily suppressed at all viewing angles due to physical emptying of the disk. 
This together with the subsequent refilling process may observationally appear as an asymmetric U-shaped dip in the soft X-ray blackbody component, accompanied by transient quasi-periodic oscillatory (QPO) signals predominantly in the EUV and Soft X-ray bands, as presented in our synthetic light curves in Figure \ref{inner}(a) and (b).
Specifically, the duration of suppression in the $22.5^\circ$ case ($\sim80$ days) is significantly longer than that in the $45^\circ$ case ($\sim$ 20 days).
Due to the enhanced density of the inner disk refilled by stellar debris and the long-lived persistent spiral arms, the emission from the inner disk is expected to increase and eventually exceed its initial value after the dip, as shown in Figure \ref{inner} (b). 
This behavior is similar to the prediction for an AGN undergoing a prograde TDE, as proposed by \cite{McKernan_2022}.

In addition, oscillatory behavior of the accretion flow with a progressively shortening period following the opening of the cavity is clearly identified in both model R22.5 and model R45, as shown in Figures \ref{QPO_snapshot} and \ref{QPO}.
These oscillations, absent in the Newtonian model N45, are driven by GR apsidal precession of the fluid near the cavity edge. 
This process quasi-periodically enhances the dissipation of orbital energy, facilitates faster infall of debris accumulated just outside the cavity, and thereby accelerates its refilling on a timescale shorter than that of viscous evolution.

In contrast, for the tilted disk scenario (the perpendicular and projected retrograde cases, models R90 and R135), as fallback material settles, the newly formed inner disk behaves like a circularized TDE disk with faster inflow and higher density compared to the previous AGN disk. 
As a result, an enhanced blackbody component is expected to appear from the extreme ultraviolet (EUV) to the soft X-ray band. 
Notably, during the fallback process, the inclination of the inner disk continuously evolves. 
If the inner disk's equatorial plane happens to sweep across the observer's line of sight, a temporary edge-on phase will occur. 
This will also lead to a dip, but with a symmetric V-shaped pattern in the UV/soft X-ray light curves, visible only from specific viewing angles (see Figure \ref{inner} (c) and (d)).
However, if the inner disk does not sweep across the observer's line of sight at any time during its evolution, the observer will instead detect a gradually enhanced UV/soft X-ray emission. 

Although both the above two types of inner disk evolution may result in a temporary disappearance of inner disk emission, two key features can distinguish them. 
First, compared to the symmetric V-shaped dip caused by the viewing angle change of a gradually tilting inner disk, the U-shaped dip related to the formation of a cavity features more temporally extended suppression near its minimum. 
Second, the U-shaped dip is accompanied by quasi-periodic oscillations, which are not observed in the V-shaped dip.

\section{Discussions}\label{sec:Discussions}
\subsection{Applications to AT2021aeuk}\label{sec:comparison_to_obs}
We applied the synthetic light curves in Figure \ref{precursor} and Figure \ref{inner} to provide implications for the multi-band observations of an AGN TDE candidate, AT2021aeuk, involving an SMBH with $M_{\rm BH}\sim10^7M_\odot$ (e.g., \citealt{Bao_2024}, \citealt{2025ApJ...982..150S}), an order of magnitude higher than in our simulations.
Since lower mass SMBHs have hotter disks at the same scaled radius (in units of $R_g$), the corresponding emission shifts to shorter wavelengths as $M_{\rm BH}$ decreases.
According to the standard thin disk solution \citep{1973A&A....24..337S}, the radial profile of blackbody effective temperature is
\begin{equation}\begin{aligned}
T_{\rm eff}(R) &\approx 1.2\times10^6\text{K}\cdot\\
& \left(\frac{M_{\rm BH}}{10^7M_\odot}\right)^{-1/4}\left(\frac{\dot{M}}{0.2\dot{M}_{\rm Edd}}\right)^{1/4} \left(\frac{R}{R_g}\right)^{-3/4},
\label{Eq_SSD_temp}
\end{aligned}\end{equation}
where $\dot{M}$ is the accretion rate and $\dot{M}_{\rm Edd}$ the Eddington accretion rate.
Additionally, because our simulation does not include radiation pressure, the effective temperature derived from local energy dissipation (see Appendix \ref{sec:spec_calc} for details) is systematically higher than that predicted by Equation \ref{Eq_SSD_temp}.
To maintain a consistent and meaningful comparison, we do not attempt to directly match the results of our simulations to the standard disk.
Instead, we use Equation \ref{Eq_SSD_temp} with $M_{\rm BH}=10^7M_\odot$ and $\dot{M}=0.2\dot{M}_{\rm Edd}$ as a reference to map the observed multi-wavelength variability to characteristic scaled disk radii based on their effective temperatures, and compare with the synthetic light curves at the corresponding scaled radii in our simulations.

\subsubsection{The precursor flare as a distinctive feature of AGN TDEs}
A precursor flare was reported $\sim200$ days prior to the onset of the first major flare in the optical light curve of AT2021aeuk (\citealt{Bao_2024}, \citealt{2025ApJ...982..150S}).
Although other possible origins of this precursor flare cannot be ruled out, if it indeed resulted from a strong collision between the star and the disk following the first pericenter passage, it would mark the actual disruption time and provide a potential diagnostic for distinguishing TDEs from intrinsic AGN variability.

According to Equation \ref{Eq_SSD_temp}, the optical light curves of AT2021aeuk correspond to effective scaled radii of $\sim700$-$1500R_g$ for g-band and $\sim1100$-$2000R_g$ for r-band, which in our synthetic light curves map to photon energy bands of roughly $6$eV and $9$eV, respectively. These central photon energies fall in the Near-UV (NUV) and Far-UV (FUV) ranges, respectively.
Therefore, we compare the observed r- and g-band optical variability in \citet{Bao_2024} and \citet{2025ApJ...982..150S} with the simulated band-integrated flux in the 2-7eV and 7-12eV bands, respectively (Figure \ref{precursor}).

Under the assumption of a TDE origin for the precursor flare in AT2021aeuk, we estimate two quantities from our simulations: the duration ratio between the major and the precursor flare ($T_{\rm maj.}/T_{\rm pre.}$), and their relative peak flux ratio, defined as $(F_{\rm maj.}-F_0)/(F_{\rm pre.}-F_0)$, where $F_0$ denotes the unperturbed flux.
As shown in Table \ref{table:precursor}, simulations with $\theta_{\rm inc} = 90^\circ$ and $135^\circ$ exhibit peak flux and duration ratios between the precursor and main flares that are comparable to those observed in the $g$- and $r$-band light curves of AT2021aeuk (\citealt{Bao_2024}, \citealt{2025ApJ...982..150S}), suggesting that the precursor flare may result from early debris-disk interaction. 
The prominent g-band precursor flare of AT2021aeuk supports the scenario of an AGN TDE at a significant orbital inclination. 

A range of precursor features have been theoretically proposed and observationally suggested in diverse TDE systems. For instance, for a white dwarf disrupted by an IMBH, extreme compression during its first pericenter passage may trigger nuclear reactions, producing additional signals prior to the TDE flare (e.g., \citealt{Luminet_Pichon+1989}; \citealt{Rosswog+2009}; \citealt{2017ApJ...839...81T}). 
Although TDEs of main-sequence stars are unlikely to produce observable signatures before fallback, transient X-ray emission during the optical rise has nonetheless been detected \citep{Malyali_2024}, possibly driven by pericenter shocks at the onset of fallback \citep{Steinberg_Stone_2024}.
An important exception occurs in AGN environments, where the star inevitably undergoes a violent collision with the disk during disruption, potentially generating observable precursor emission.
Whether precursor flares are a common feature in AGN TDEs remains an open question. The discovery of additional precursor flares from AGN TDEs in future wide-field time-domain surveys will be crucial for drawing statistically robust conclusions.
\begin{table*}[ht]
\scriptsize
\tabcolsep=0.4cm
\caption{
Comparison of the peak flux ratio and duration ratio between the major and the precursor flare from our models and optical observations of AT2021aeuk.
\label{table:precursor}}
\centering
\begin{tabular}{ccccc}
\hline
& Energy Band & Effective Radius from& Peak Flux Ratio & Duration Ratio \\
&             &  Eq. \ref{Eq_SSD_temp} (observations) or Fig. \ref{init_spec_teff} (simulations)& ($\frac{F_{\rm maj.}-F_0}{F_{\rm pre.}-F_0}$) & ($T_{\rm maj.}/T_{\rm pre.}$) \\
\hline\hline
\multicolumn{5}{c}{Simulations}\\
model &  &  &  &\\
\hline
\multirow{2}{*}{ R22.5}
 & 2-7 eV  (Optical/NUV) & $>1500R_g$ & $\sim$1.8  & $>15$ \\
 & 7-12 eV  (FUV)        & $\sim700-1500R_g$   & $\sim$14   & $>15$ \\
\multirow{2}{*}{R45}
 & 2-7 eV   & $>1500R_g$  & $\sim$1.8  & $>15$ \\
 & 7-12 eV          & $\sim700-1500R_g$  & $\sim$20   & $>15$ \\
\multirow{2}{*}{R90}
 & 2-7 eV   & $>1500R_g$  & $\sim$1.4  & $\sim$9.0 \\
 & 7-12 eV       & $\sim700-1500R_g$  & $\sim$6.0  & $\sim$9.0 \\
\multirow{2}{*}{R135}
 & 2-7 eV   & $>1500R_g$  & $\sim$2.2  & $\sim$9.0 \\
 & 7-12 eV      & $\sim700-1500R_g$  & $\sim$5.6  & $\sim$9.0 \\
\hline\hline
\multicolumn{5}{c}{Observations}\\
\hline
\multirow{2}{*}{AT2021aeuk}
 & combined light curve[1] & $\sim700-2000R_g$  & $\sim$3.2 & $\sim$6.5 \\
  & g-band, fitted[2] & $\sim700-1500R_g$  & $\sim$5.0 & $\sim$6.5 \\
\hline
\end{tabular}
\tablecomments{
[1] The combination of photometry data of ZTF, ASAS-SN, ATLAS, CRTS, and PS1 (\citealt{Bao_2024}, Fig. 2).\ \ \ \ \ \ \ \ \ \ \ 
[2] The fitted g-band light curve of ZTF and the Lijiang 2.4m telescope (\citealt{2025ApJ...982..150S}, Fig. 4).}
\end{table*}

\subsubsection{The UV dip indicative of inner disk evolution}
A prominent late-time UV dip following the UV flare is detected in the Swift UVW1-band accompanied by the second optical flare of AT2021aeuk, as shown in Figure 8 of \citet{2025ApJ...982..150S}, suggesting a temporary suppression of the inner disk radiation.

It is important to note that AT2021aeuk is likely a partial TDE \citep{2025ApJ...982..150S}. 
In this scenario, the perturbed radius scaled by $R_g$ is relatively large because partial TDEs typically occur at larger pericenter distances compared to full disruptions (e.g., \citealt{Ryu_2023_ApJ}).
As a result, the dominant emission for the dip shifts to longer wavelengths compared to fully disrupted cases, such as those in our models.
Therefore, when comparing our simulations with the UV light curve of AT2021aeuk, it is inappropriate to simply match the disk radii by rescaling; instead, we should assume that the effective radius of the UV dip corresponds roughly to the TDE's perturbed radius.
In our simulation, the perturbed radius is $\sim50-100R_g$, where the local emission is dominated by EUV band and extended to soft X-ray in the synthetic light curves.
Dip features similar to the UV variability in AT2021aeuk can be seen in these bands, as shown in Figure \ref{inner}. 

According to our simulations, the UV dip in AT2021aeuk can be attributed either to a geometric effect, in which the inner disk gradually tilts and temporarily aligns edge-on with the observer (as in models R90 and R135), or to a physical depletion of the innermost region, which is subsequently refilled (as in model R45).
We suggest that the geometric effect of a tilted inner disk as in our model R90 and R135 is more favorable as the flux after the dip does not exceed the unperturbed level.
However, more substantial evidence is still required for observationally distinguishing them.

Based on our theoretical predictions, we expect that long-term monitoring of AGNs will reveal more AGN TDE candidates with dips from UV to soft X-ray, corresponding to recovery flux that is lower than, higher than, or comparable to the initial flux, and whether or not accompanied by quasi-periodic oscillations, thus allowing observational verification of these two mechanisms.

\subsection{Implications for The X-ray Corona}
Although we did not dynamically evolve the X-ray corona, our model can provide insights into the disappearance of the X-ray power-law component (e.g., \citealt{2019ApJ...883...94T}, \citealt{Ricci_2020}, \citealt{2025ApJ...982..150S}).

Previous studies suggest that a hot, ionized gas corona is attached to the surface of the inner region of the AGN accretion disk (e.g., \citealt{1991ApJ...380L..51H}, \citealt{1993ApJ...413..507H}, \citealt{2002ApJ...575..117L}). 
The soft photons generated by the disk's blackbody radiation are Comptonized by the corona, producing the X-ray power-law component. 
The disk magnetic field is believed to play a key role in powering the corona (e.g., \citealt{Merloni_Fabian_2001}, \citealt{Liu_2016}, \citealt{2020MNRAS.495.1158C}).
For a corona coexisting with the disk, changes in the disk state during the fallback phase of a TDE are expected to strongly affect the corona.
This may occur either by cutting off the power supply and terminating the inflow of soft photons \citep{Ricci_2020}, or by directly disrupting the coronal structure through strong outflows triggered by the TDE \citep{Cao_2023}.
If the inner accretion disk is depleted (as in our models R22.5 and R45), the disruption of magnetic field and the suppression of seed photons can cause the X-ray power-law component to disappear, leaving only a soft thermal continuum. As the fallback stream weakens and the cavity refills, the hard X-ray emission gradually recovers.
In the case of a misaligned disk (as in our models R90 and R135), the tilted inner regions can intrude into the original coronal area and alter the magnetic field configuration.    
Such a perturbation on the corona may also cause the X-ray power-law component to disappear temporarily until the disk settles.
The detailed suppression and recovery process of the X-ray power-law component require careful dynamical modeling and proper radiative transfer treatment of the corona, warranting further in-depth investigation.

\subsection{Implications for QPOs and QPEs}
In the weak-field limit of the Schwarzschild solution, the apsidal precession angle per orbit can be approximated as (e.g., \citealt{1972gcpa.book.....W})
\begin{equation}
    \Delta\phi = \frac{6\pi GM_{\rm BH}}{a(1-e^2)c^2},
\end{equation}
where $a$ and $e$ are the semi-major axis and eccentricity of the orbiting particle.  
In our simulations, where the typical radius of the central cavity is $\sim 100R_g$, this approximation remains accurate to describe the precessional behavior of the fluid on the cavity edge.  
The corresponding precession period of the cavity edge is
\begin{equation}
    \label{Eq_prec_period}
    T_{\rm prec} \approx 10.9\ \text{days} \left(\frac{M_{\rm BH}}{10^6M_\odot}\right)\left(\frac{a_{\rm cav}}{100R_g}\right)^{5/2}(1-e_{\rm cav}^2),
\end{equation}
where $a_{\rm cav}$ and $e_{\rm cav}$ denote the semi-major axis and the eccentricity of the cavity, respectively.

\begin{figure*}[ht]
\centering
\includegraphics[width=0.9\linewidth]{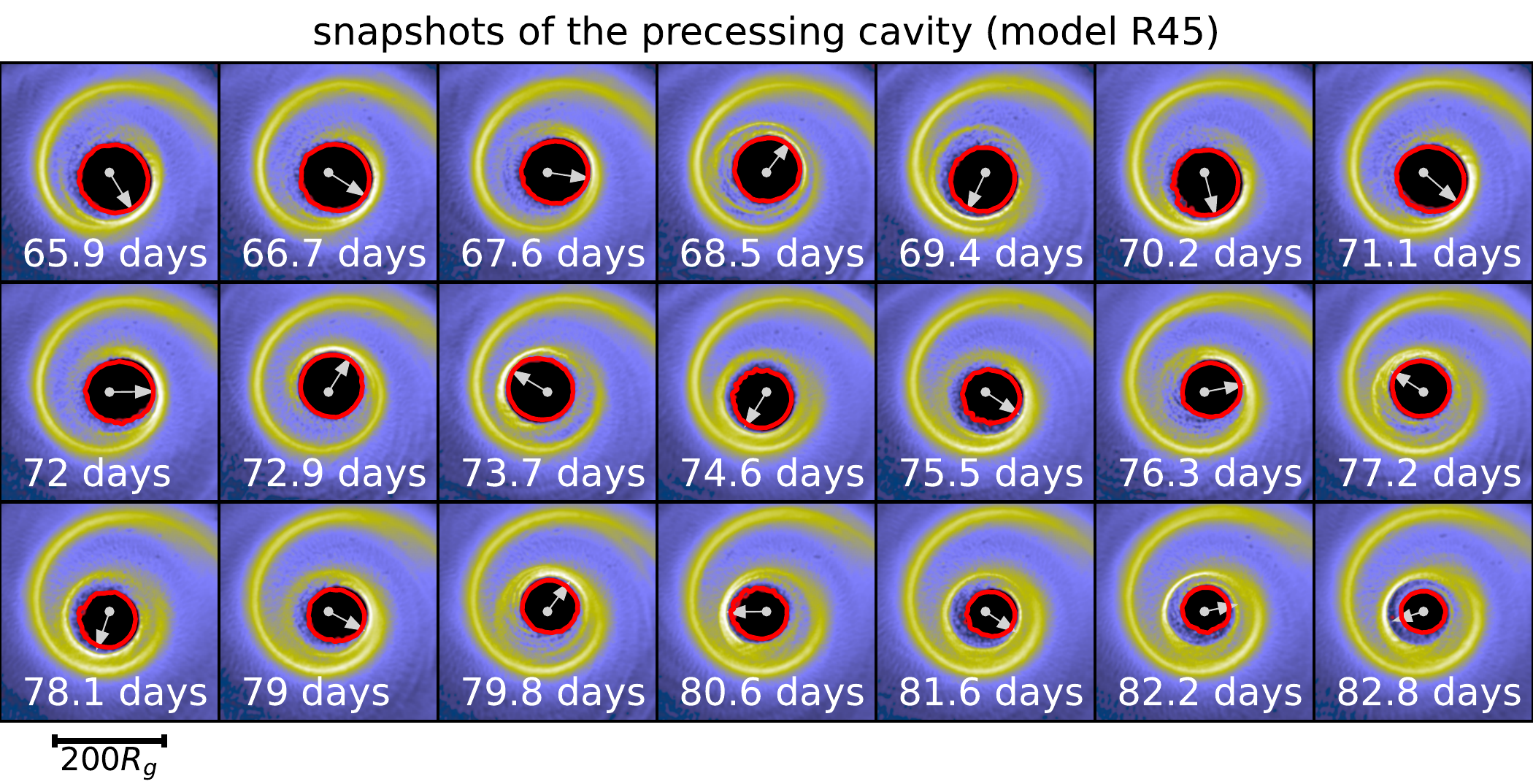} 
\caption{\label{QPO_snapshot}
Snapshots of the central cavity for model R45.
In each panel, the red curve outlines the cavity edge, the gray dot indicates the central SMBH, and the gray arrow denotes the direction of the cavity's major axis.
The time since the beginning of the simulation is labeled in each panel.
A scale bar of $200R_g$ is shown in the bottom left of the plot.
}
\end{figure*}

We identify prominent quasi-periodic signals in the inner disk radiation (EUV/soft X-ray) during the dip phase in models R22.5 and R45. As shown in Figure \ref{inner}, this variability is characterized by oscillations with decreasing period, occurring between 100 and 160 days in model R22.5 and between 65 and 83 days in model R45. 
These signals are closely correlated with the precession of the central cavity, whose size gradually decreases over time (see Figure \ref{QPO_snapshot}). 
The precession induces structural oscillations in the cavity, leading to repeated orbital energy dissipation near the cavity edge and powering the observed quasi-periodic emission in the EUV/soft X-ray bands. 

\begin{figure*}[ht]
\centering
\includegraphics[width=0.85\linewidth]{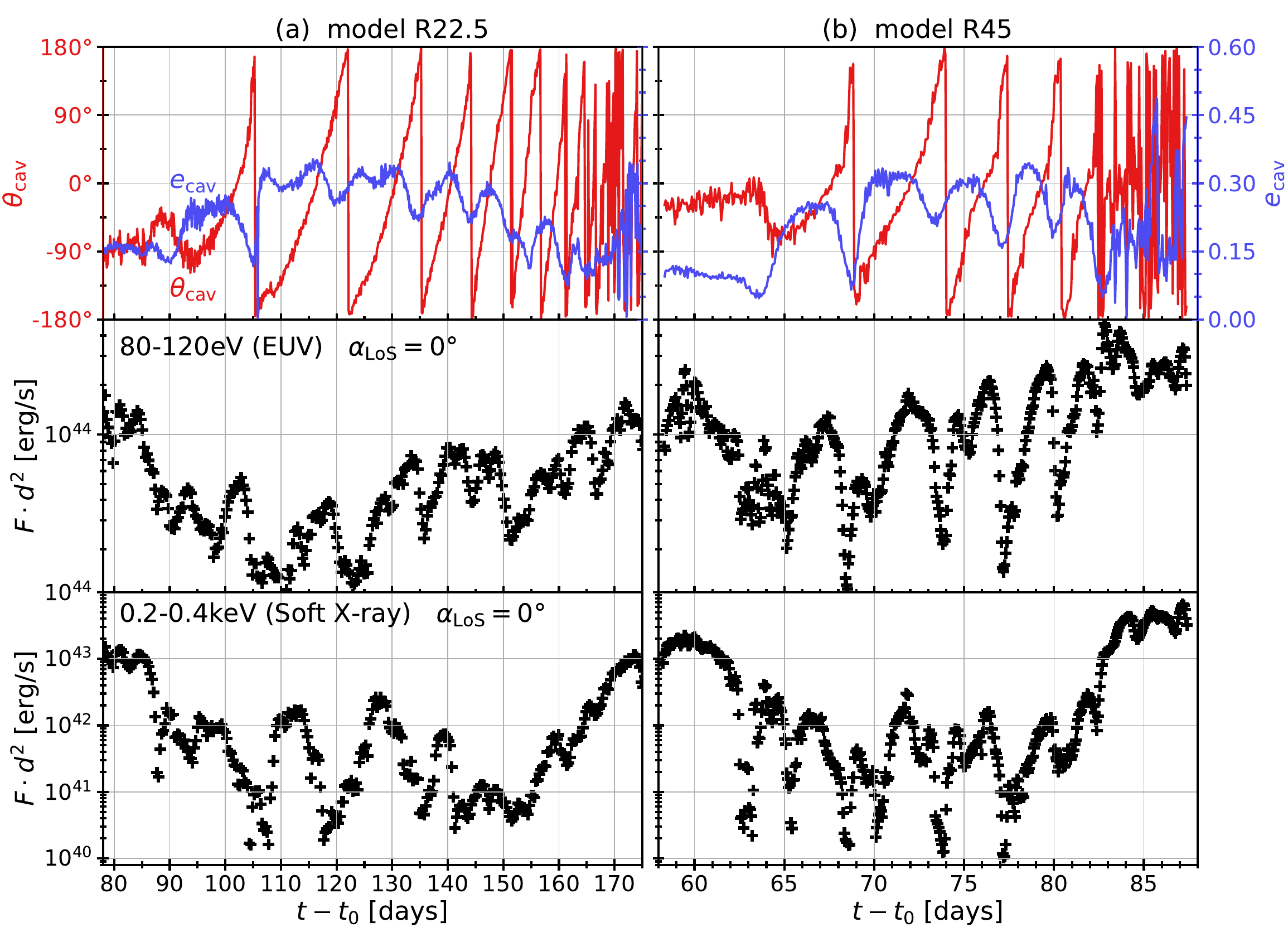} 
\caption{\label{QPO}
Comparison of cavity precession and inner disk light curve for (a) model R22.5 and (b) model R45.
Top panels show the cavity's eccentricity ($e_{\rm cav}$, blue) and major-axis orientation ($\theta_{\rm cav}$, red) as a function of time during its precession. 
Pronounced oscillations occur from $\sim$ 100 to 160 days in model R22.5 and from $\sim$ 65 to 83 days in model R45, characterized by the rotation of the cavity's major axis and quasi-periodic variations in its eccentricity, temporally aligned with the oscillation signals in EUV and Soft X-ray light curves shown in the middle and bottom panels.  
}
\end{figure*}

Figure \ref{QPO} further quantifies this behavior, showing the time evolution of the cavity's eccentricity ($e_{\rm cav}$) and major-axis orientation ($\theta_{\rm cav}$). Both quantities show five distinct cycles with an average period of $\sim3.6$ days in model R45, and about seven cycles with an average period of $\sim10$ days in model R22.5. 
Assuming a cavity eccentricity $e_{\rm cav} \approx 0.3$ and using Equation~\ref{Eq_prec_period}, we derived a precession period of $\sim9.92$ days for model R22.5 ($a_{\rm cav} \approx 100R_g$) and $\sim4.07$ days for model R45 ($a_{\rm cav} \approx 70R_g$), in good agreement with the numerical results. 
We note that these QPO signals may span a range of frequencies, depending on both $M_{\rm BH}$ and the size of the oscillating fluid structure. Equation \ref{Eq_prec_period} thus provides a useful tool for identifying QPOs driven by the precession of the cavity induced by TDEs in AGNs, and may serve to constrain the size of precessing cavities in systems with known black hole masses.

Moreover, Lense-Thirring effect from a spinning SMBH with disk misaligned with its spin may provide an alternative mechanism for generating QPO signals due to both dynamical effect and viewing angle dependence (e.g., \citealt{Stella_1998}, \citealt{2019MNRAS.487..550L}).
Since misaligned inner disks can be generated in AGN TDEs, as demonstrated in our perpendicular and projected retrograde cases (models R90 and R135), we expect that their misalignment with the SMBH spin may induce additional quasi-periodic motion driven by the Lense-Thirring effect during the realignment process.
Detailed dynamical features and their potential connection to QPOs in AGN TDEs involving spinning SMBHs are beyond the scope of our current setup and will be explored in a following study. 

While the QPO signals from our models R22.5 and R45 arise directly from our simulations, only indirect insights can be offered into the potential connection between AGN TDEs and QPEs within our setup. 
We note that our high-inclination simulations (models R90 and R135) naturally produce a misaligned inner disk, which is a feature commonly invoked in models for QPEs (e.g., \citealt{Zhou+2024}, \citealt{Raj_2021a}, \citealt{Raj_2021b}). 
Since stars disrupted in AGN disks can have arbitrary inclinations \citep{2024ApJ...962L...7W}, the resulting angular momentum inheritance provides a natural pathway for forming such misaligned disks, thereby potentially facilitating the conditions required for QPEs. 
One proposed mechanism involving a misaligned disk is the Extreme Mass Ratio Inspiral interacting with a TDE disk \citep{Zhou+2024}, in which a stellar-mass object migrates inward through the AGN disk, orbits an SMBH and repeatedly collides with two distinct regions of a surrounding misaligned TDE disk, naturally producing alternating long-short recurrence intervals and strong-weak intensities, as observed in some QPE sources. 
Another proposed mechanism involves the tearing of a warped, misaligned accretion disk, where the Lense-Thirring effect induces quasi-periodic enhancements in the accretion rate through repeated misalignment and realignment of inner disk segments (\citealt{Raj_2021a}, \citealt{Raj_2021b}).

\section{Conclusion}
Building upon the foundational work of \cite{2019ApJ...881..113C} and \cite{2024MNRAS.527.8103R}, we systematically investigate the impact of stellar orbital inclination ($\theta_{\rm inc}$) and GR effects on the evolution of AGN TDEs. 
Using the Godunov-Type Lagrangian hydrodynamic code GIZMO \citep{2015MNRAS.450...53H}, we conduct a series of simulations that incorporate both Newtonian and GR gravitational potentials. 
In contrast to previous studies, our models include an optically thick blackbody radiative cooling term (Section \ref{sec:rad_cooling}) and assume a highly eccentric bound stellar orbit (Section \ref{sec:orbit}).

Key results are as follows:\\

(1) \textbf{Orbital inclination strongly influences inner disk morphology via angular momentum transfer (Section \ref{subsec:Inc}, Figure \ref{innerdisk})}.
Prograde encounters form a central cavity with spiral arms, while perpendicular and retrograde encounters produce a misaligned, debris-dominated inner disk. 
These morphological differences arise from distinct angular momentum transfer mechanisms and result in significantly different inflow rates (Figure \ref{AM_and_inflow_rate}).

(2) \textbf{
GR effects alter debris trajectories and the effective potential experienced by the disk, leading to substantial changes in the structures of both the fallback material and the disk (Section \ref{sec:GR}).
}
Precession-induced orbital deviations shorten the debris–disk interaction phase and reshape both fallback streams and the outer disk (Figure \ref{GR1}).
Together with the deeper potential well in the innermost region, these effects further give rise to distinct inner disk morphologies compared to the Newtonian case (Figure \ref{Sch_Kep_inner_disk}).

(3) \textbf{Our simulations produce a two-phase UV/optical light curve, consisting of a precursor flare from early debris–disk interaction, followed by a major fallback flare, with both components sensitive to $\theta_{\rm inc}$.}
For a projected retrograde orbit of the star, a more significant precursor flare is observed, with the major flare exhibiting a sharper rise and fall, a peak bolometric luminosity several times higher, and a duration shorter than in the projected prograde case (Section \ref{sec:variability}, Figure \ref{precursor}).

(4) \textbf{Two distinct mechanisms for dips of inner disk radiation accompanying the UV/optical fallback flare are revealed from our simulations: physical clearing of the inner disk in projected prograde cases, and geometric obscuration by a tilted inner disk in retrograde and perpendicular cases (Section \ref{sec:innerdisk_lc}, Figure \ref{inner}).} 
In the projected prograde cases where a central cavity forms, the radiation from the innermost disk exhibits a U-shaped dip accompanied by quasi-periodic oscillations, 
associated with the emptying and refilling of the inner disk, with the latter process being expedited by the apsidal precession of the cavity edge. 
In contrast, for the tilted disk scenario in the perpendicular and projected retrograde cases, 
a symmetric V-shaped dip will be observed at specific viewing angles if the equatorial plane of the tilted disk happens to sweep across the line of sight.

(5) \textbf{A UV/Optical precursor flare followed by a late-time dip in inner disk radiation may serve as distinguishing signatures of AGN TDEs, reflecting early debris-disk interaction and subsequent disruption or tilt of the inner accretion flow (Section \ref{sec:comparison_to_obs}).}
The prominent precursor flare and late-time dip in inner disk radiation observed in AT2021aeuk (e.g., \citealt{2025ApJ...982..150S}) are consistent with the expected signatures of an AGN TDE involving a highly inclined stellar orbit.

(6) \textbf{Our simulations suggest that AGN TDEs can give rise to quasi-periodic signals, with distinct origins depending on the orbital inclination of the disrupted star.}
In AGN TDEs with a projected prograde stellar orbit, QPOs can arise from Schwarzschild apsidal precession of the fluid near the central cavity carved out by the returning debris stream. 
The oscillation timescale is governed by the black hole mass ($M_{\rm BH}$), as well as the size and eccentricity of the precessing cavity (Equation \ref{Eq_prec_period}).
In contrast, the inclined inner disks formed in perpendicular and projected retrograde AGN TDEs may create favorable conditions for the emergence of QPEs, possibly through repeated collisions of a stellar-mass object with the tilted inner disk, or through episodic accretion events triggered by disk misalignment.

We anticipate that future advances in time-domain astronomy will make it possible to observationally confirm the phenomena of AGN TDEs proposed in this study, such as precursor flares, UV/soft X-ray dips, and quasi-periodic signals, via long-term multi-wavelength monitoring of large AGN samples.

\begin{acknowledgments}
M.Z. and W.Z. would like to acknowledge the support by the National Natural Science Foundation of China (Grant No. 12333004, 12433005), and the support by the Strategic Priority Research Program of the Chinese Academy of Sciences, Grant No. XDB0550200.
H.G. acknowledges the National Natural Science Foundation of China (NFSC, No. 12473018)  and Overseas Center Platform Projects, CAS, No.  178GJHZ2023184MI.
M.Z. would also like to thank Xin Pan, Dongyue Li, Mingjun Liu and Patrik Mil\'an Veres for helpful discussions.
The authors acknowledge the Beijing Super Cloud Computing Center (BSCC) and the National Supercomputing Center, Jinan for providing high-performance computing resources that have contributed to the research results reported within this paper. 
We also sincerely appreciate the prompt and professional support from their technical teams.
\end{acknowledgments}

\software{GIZMO \citep{2015MNRAS.450...53H}, 
SPLASH \citep{2007PASA...24..159P},
Python, 
NumPy,
SciPy,
h5py, 
Matplotlib, 
FFmpeg}

\appendix
\section{Resolution Tests}
In TDE simulations performed with the SPH method, the accuracy of structure formation and evolution is highly sensitive to numerical resolution (e.g., \citealt{Liptai_2019}, \citealt{Bonnerot_2022}, \citealt{2024ApJ...971L..46P}). 
A well-known example is the stream-fanning phenomenon after nozzle shock observed in vacuum TDE simulations, which is highly resolution dependent and remains a numerical challenge \citep{2024ApJ...971L..46P}. 
In our AGN TDE simulations, stream fanning does not occur, as the fallback debris promptly interacts with the pre-existing AGN disk after the nozzle shock and rapidly merges into it. 
This interaction suppresses stream-fanning and instead leads to disk perturbations that induce substructures such as a central cavity and spiral arms (e.g., Figure \ref{innerdisk}). 

To assess the numerical convergence of our simulation results, we performed a resolution test for an orbital inclination of $\theta_{\rm inc} = 45^\circ$ using three different total particle numbers:
$N_{\rm particle} = 3.5 \times 10^5$, $7.0 \times 10^5$, and $1.4 \times 10^6$, with corresponding disk and stellar particle allocations of
$(N_{\rm disk}, N_{\rm star}) = (2.5 \times 10^5, 1.0 \times 10^5)$, $(5.0 \times 10^5, 2.0 \times 10^5)$,
and $(1.0 \times 10^6, 4.0 \times 10^5)$, respectively.
To further examine convergence under different inner disk morphologies, we also performed an additional resolution test for $\theta_{\rm inc} = 90^\circ$ using two total particle numbers: $N_{\rm particle} = 7.0 \times 10^5$ and $1.4 \times 10^6$, with the same disk and stellar particle allocations as in the $\theta_{\rm inc} = 45^\circ$ case.

\begin{figure*}[ht]
\centering
\includegraphics[width=0.8\linewidth]{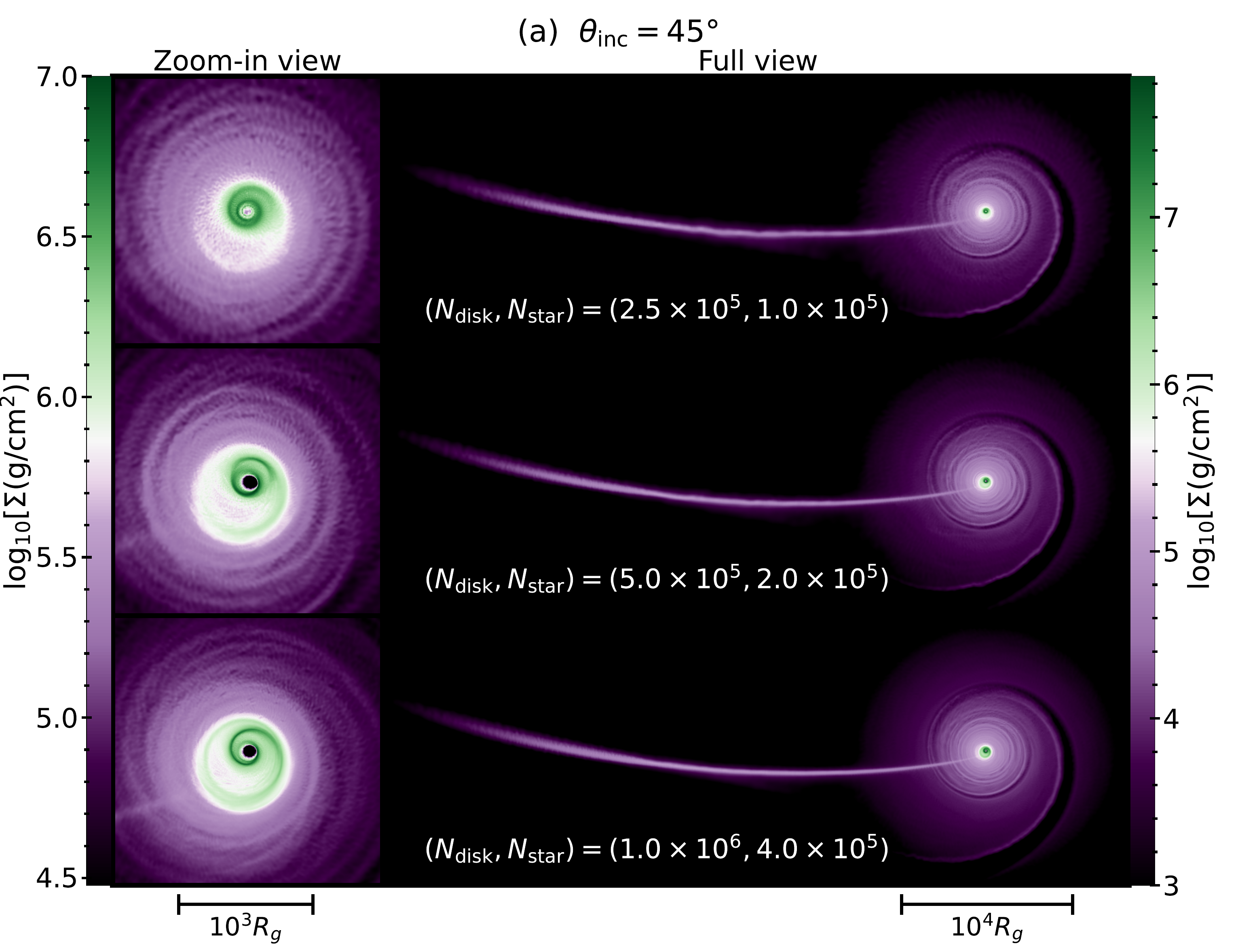} 
\includegraphics[width=0.8\linewidth]{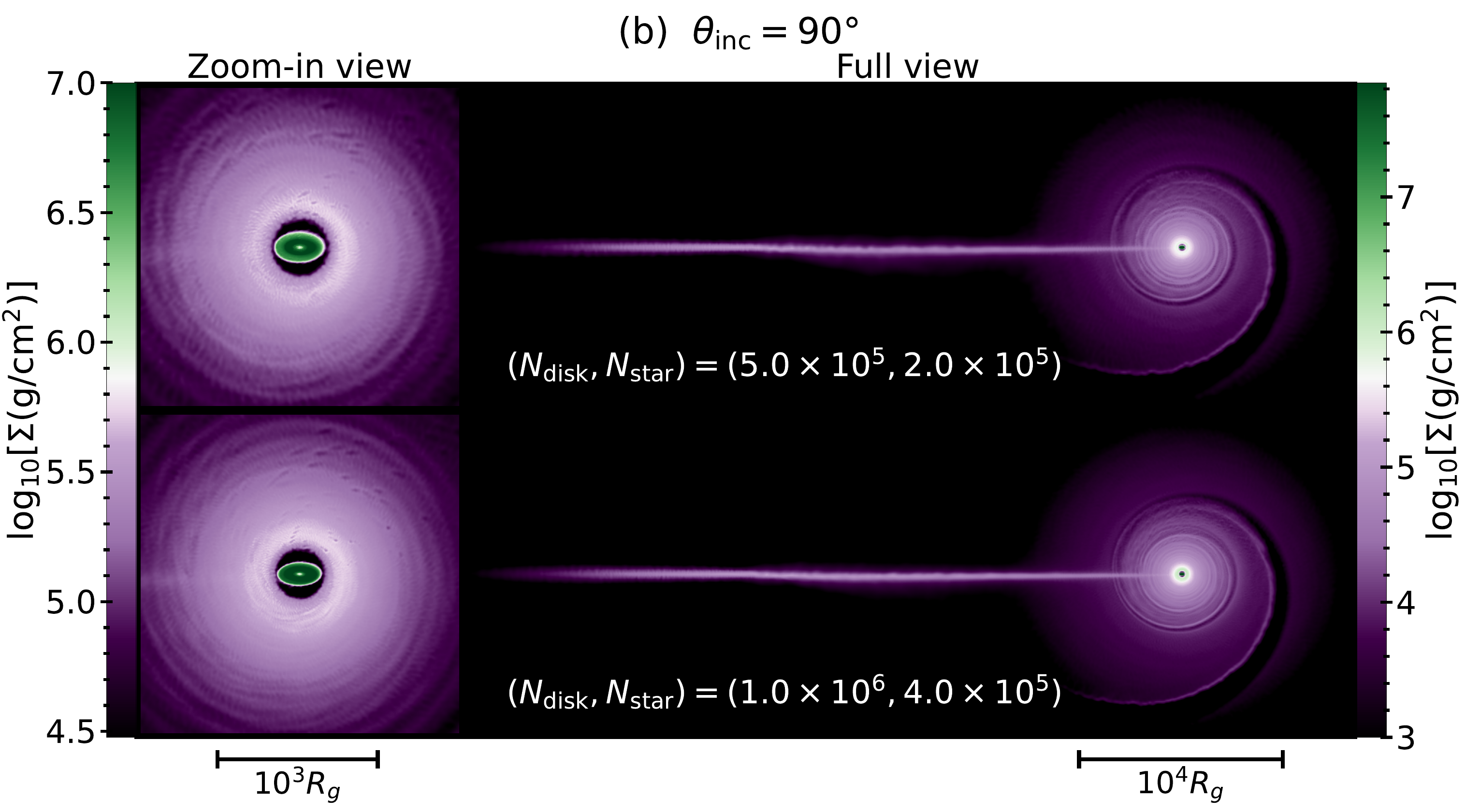} 
\caption{\label{RT}
Face-on Surface density maps from the resolution test at $t = 75.8$ days are shown for two cases: (a) orbital inclination $\theta_{\rm inc} = 45^\circ$, and (b) $\theta_{\rm inc} = 90^\circ$.
The number of particles for both the disk ($N_{\rm disk}$) and the star ($N_{\rm star}$) is indicated in each panel.
The left column provides a zoom-in view of the inner disk region.
The right column displays the global structure, including the debris stream and the perturbed disk.
}
\end{figure*}

Surface density maps at a well-evolved stage ($t = 75.8$ days) for both inclination angles and various resolutions are presented in Figure \ref{RT}. 
We find that the key morphological features, such as the disrupted stellar debris and the large-scale response of the outer disk, are reasonably well resolved at $N_{\rm particle} = 7.0 \times 10^5$. 
However, finer structures in the inner regions show more noticeable sensitivity to resolution. For instance, lower-resolution runs tend to exhibit faster cavity refilling and less distinct spiral patterns in the projected prograde cases, likely due to enhanced numerical dissipation.
To ensure accurate modeling of small-scale features in the inner region and to minimize numerical artifacts, we adopt the highest resolution, $N_{\rm particle} = 1.4 \times 10^6$, as our fiducial setup in the regular runs.

\section{Estimations of Radiative Features}\label{sec:radiative_features}
Since our model does not explicitly account for radiative transfer and include radiation pressure in the EOS, we cannot accurately track the temperature of the accretion flow and thus cannot directly compute its radiation from the simulated temperature. 
However, under the assumption of local thermodynamic equilibrium (LTE), we can relate energy dissipation to blackbody radiation to obtain the effective temperature ($T_{\rm eff}$) of the disk.
In this section, we introduce our approach for estimating bolometric luminosity and disk spectrum based on the energy dissipation rate. 

\subsection{Calculation of Total Energy}\label{sec:energy_loss_calc}
The orbital energy
$
E_{\rm orbit}=\sum_i \frac{1}{2}m_iv_i^2-\frac{GM_{\rm BH}m_i}{R_i}
$
and the internal energy 
$
E_{\text{I}}=\sum_i m_iu_i
$
of a given area are calculated at each time step by summing over all particles $i$ in the area, with the total energy 
\begin{equation}
E_{\rm tot} =  E_{\rm orbit}+E_{\rm I}.
\end{equation}
Especially, for our GR models, we adopt the orbital energy corresponding to the relativistic potential
\begin{equation}\begin{aligned}	
E_{\rm orbit,R}=
\sum_i&\frac{1}{2}\left(\frac{R_i}{R_i-2R_g}\right)^2m_iv_{r,i}^2+\frac{1}{2}\left(\frac{R_i}{R_i-2R_g}\right)m_iv_{t,i}^2
-\frac{GM_{\rm BH}m_i}{R_i}
\end{aligned}\end{equation}
as in \cite{2016MNRAS.455.2253B} to include the relativistic correction in calculating orbital energy.

\subsection{Calculation of Effective Temperature and Disk Spectrum}\label{sec:spec_calc}
According to energy conservation, the total energy loss in a certain region mainly comes from two parts: radiative cooling and advection. 
This can be written as
\begin{equation}
\frac{\partial e_{\rm tot}}{\partial t}+\nabla\cdot \textbf{\emph{S}} = \Lambda_{\rm cool},
\end{equation}
where $e_{\rm tot}$ is the total energy density, $\textbf{\emph{S}}$ is the energy flux, and $\Lambda_{\rm cool}$ is the radiative cooling rate.

To estimate the radiative energy loss of a certain region, it is important to eliminate the contribution from advective energy transport via boundary inflows and outflows.
In our Lagrangian-based simulations, by following individual fluid elements (particles), we can effectively avoid the advective term by tracking the same set of particles over time. 
To achieve this, we divide the disk into multiple narrow rings $[R_i,R_{i+1}]$, and then compute the difference in total energy between $t+dt$ and $t-dt$ with $dt\approx0.23$ days for the same set of particles (labeled $j$) that happen to be within the $j^{\rm th}$ radial bin $[R_i,R_{i+1}]$ at time $t$:
\begin{equation}
\frac{dE_{\rm tot}}{dt}|_{[R_i,R_{i+1}]}(t) = \frac{E_{\text{tot},j}(t+dt)-E_{\text{tot},j}(t-dt)}{2dt}.
\label{Eq_dEdt}
\end{equation}
By tracking the same particles, we exclude the effect of advection and directly obtain the radial profile of radiative energy loss, which is then used to estimate the local blackbody effective temperature $(T_{\rm eff}(R))$ and disk emission features.

\begin{figure*}[ht]
\centering
\includegraphics[width=0.635\linewidth]{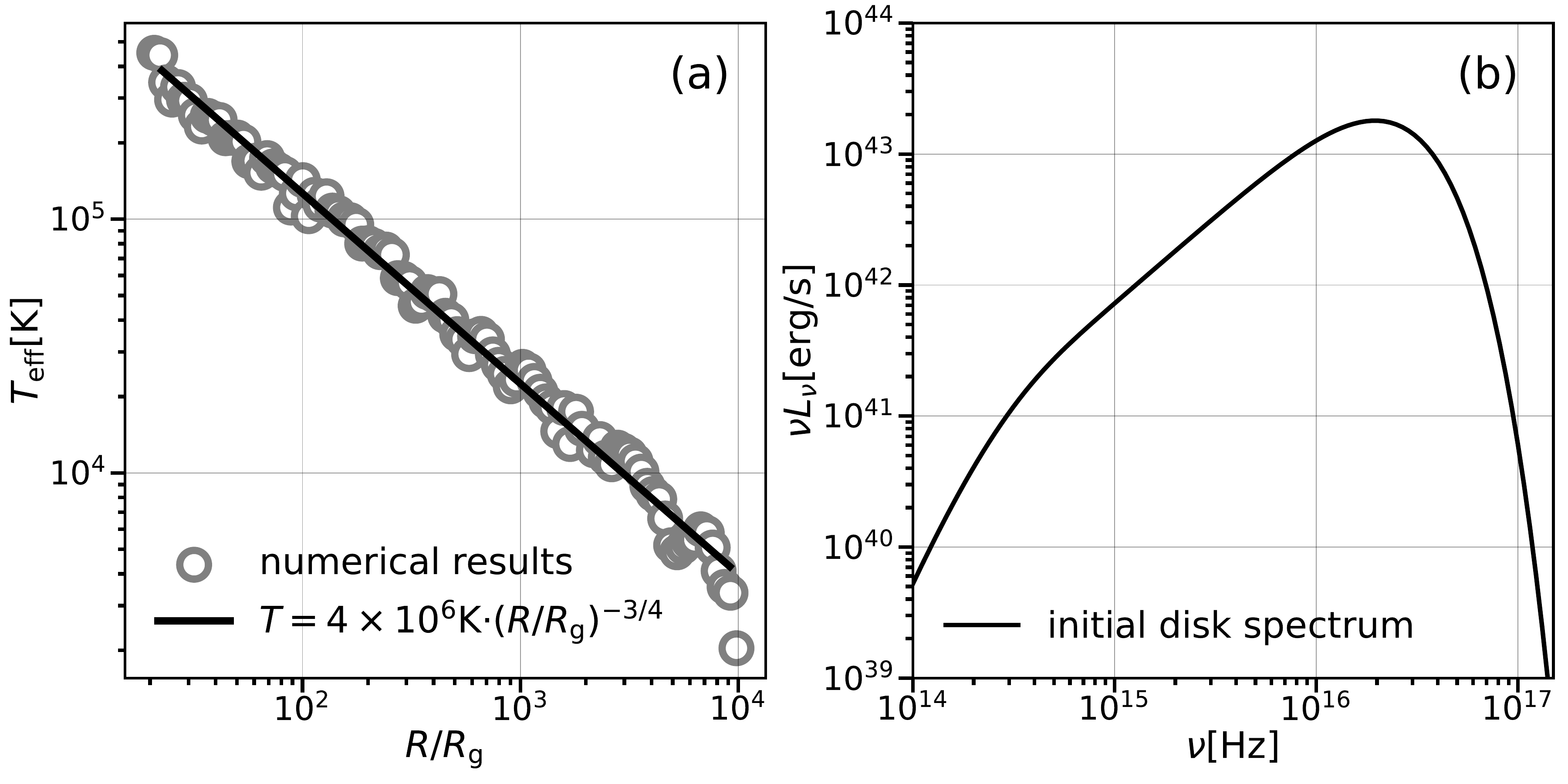}
\caption{\label{init_spec_teff} 
The left panel shows the radial profile of the blackbody effective temperature $(T_{\rm eff})$ of the initial disk, where gray circles represent the numerical results from Equation \ref{Eq_Teff}, and the black solid line is a power-law fit given by $T=4\times10^6$K$\cdot(R/R_g)^{-3/4}$. 
The right panel shows the corresponding multi-temperature blackbody spectrum ($\nu L_\nu$) of the initial disk.
}
\end{figure*}

To derive $T_{\rm eff}(R)$ of the initial disk, we equate the energy dissipation rate to the blackbody radiation power for each radial bin:
\begin{equation}
-\frac{dE_{\rm tot}}{dt}|_{[R_i,R_{i+1}]}=2(\pi R_{i+1}^2-\pi R_i^2)\sigma T_{\rm eff}(R_{j})^4,
\label{Eq_Teff}
\end{equation}
and eventually yield $T_{\rm eff}$ at $R_j=\frac{1}{2}(R_i+R_{i+1})$, as presented in Figure \ref{init_spec_teff} (a).

Once we obtain $T_{\rm eff}(R)$ of the inital AGN disk, we can calculate its multi-temperature blackbody spectrum. 
According to Planck's function, the radiative intensity per unit frequency at $R$ is
\begin{equation}
B_{\nu}[T_{\rm eff}(R)]=\frac{2h\nu^3}{c^2}\frac{1}{e^{\frac{h\nu}{k_BT_{\rm eff}(R)}}-1}.
\end{equation}
For an observer at a distance $D$ with a line of sight direction $\hat{\textbf{n}}_{\rm LoS}$, the flux at frequency $\nu$ from the initial disk can be calculated \citep{2002apa..book.....F}:
\begin{equation}\begin{aligned}
F_{\rm {init},\nu}&=\frac{\hat{\textbf{n}}_{\rm init}\cdot\hat{\textbf{n}}_{\rm LoS}}{D^2}\int_{R_{\rm in}}^{R_{\rm out}}B_{\nu}[T_{\rm eff}(R)]2\pi RdR,
\end{aligned}
\end{equation}
where $R_{\rm in}$ and $R_{\rm out}$ are the inner and outer boundary of the disk.
The numerical result of the multi-temperature blackbody spectrum is shown in Figure \ref{init_spec_teff} (b). 
\subsection{Spectrum Evolution}\label{sec:spectrum}
To estimate the evolution of the multi-temperature spectrum of the perturbed disk, we assume that the heat generated from stream collisions is immediately removed by radiative cooling, while the internal energy remains nearly unchanged. 
Under this assumption, we first compute the radial profile of energy dissipation at time $t$ $(\frac{dE}{dtdS}|_{(R,t)})$ according to Equation \ref{Eq_dEdt}. 
Next, we equate it to the power of blackbody radiation, where the blackbody temperature is taken to be the local effective temperature of the initial accretion disk ($T_{\rm eff}(R)$).
We note that in this approach the radius of the accretion disk is mapped to the azimuthally averaged effective temperature profile.

\begin{figure}[ht]
\centering
\includegraphics[width=0.54\linewidth]{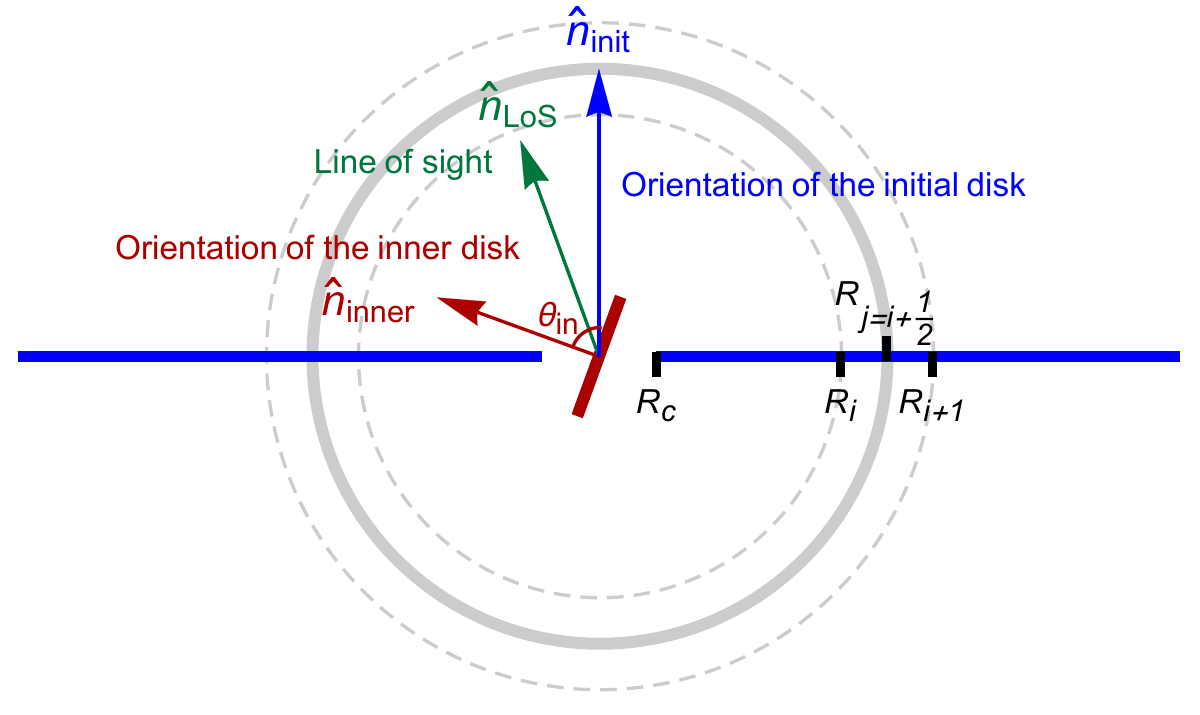}
\caption{\label{sketch} 
A sketch map that shows the definitions of the parameters to describe the configuration of the perturbed AGN disk. 
The blue region represents the initial disk with unit vector $\hat{\textbf{n}}_{\rm init}$. 
The inner disk inside $R_c$ is tilted by an angle $\theta_{\rm in}$ relative to the initial orientation after the fallback of debris stream in a misaligned orbit. Its
unit vector is $\hat{\textbf{n}}_{\rm inner}$.
The unit vector of the line of sight of the observer is $\hat{\textbf{n}}_{\rm LoS}$, marked by a green arrow.
}
\end{figure}

Distinct inner disk structures are yielded in our simulations for various orbital inclinations. 
Therefore, we consider the observational inclinations of misaligned inner and outer disks separately.
A detailed explanation of the disk configuration and orientation notations is provided in Figure \ref{sketch}.
For an observer at a distance $D$ with a line of sight direction $\hat{\textbf{n}}_{\rm LoS}$, the flux at frequency $\nu$ from the disk is
\begin{equation}\begin{aligned}
F_{t,\nu}&=
\frac{\hat{\textbf{n}}_{\rm inner}\cdot\hat{\textbf{n}}_{\rm LoS}}{D^2}\int_{R_{\rm in}}^{R_{\rm c}}\frac{dE}{dtdS}|_{(R,t)}\frac{B_\nu[T_{\rm eff}(R)]}{\sigma T_{\rm eff}^4(R)}2\pi RdR 
+\frac{\hat{\textbf{n}}_{\rm init}\cdot\hat{\textbf{n}}_{\rm LoS}}{D^2}\int_{R_{\rm c}}^{R_{\rm out}}\frac{dE}{dtdS}|_{(R,t)}\frac{B_\nu[T_{\rm eff}(R)]}{\sigma T_{\rm eff}^4(R)}2\pi RdR,
\end{aligned}
\end{equation}
where $R_{\rm c}$ is the interface between the inclined inner disk and the outer disk. 

\newpage
\bibliography{AGN_TDE_arxiv_version}{}
\bibliographystyle{aasjournal}
\newpage

\end{CJK*}
\end{document}